\documentclass[pre,superscriptaddress,amsfonts,amssymb,amsmath,letterpaper,nofootinbib,floatfix,showpacs]{revtex4}

\usepackage{graphicx}
\usepackage{bm}

\begin{document}
\newlength{\GraphicsWidth}
\setlength{\GraphicsWidth}{8cm}
\newcommand\comment[1]{\textsc{{ 1}}}
\newcommand{\be}{\begin{equation}}
\newcommand{\ee}{\end{equation}}
\newcommand{\ba}{\begin{eqnarray}}
\newcommand{\ea}{\end{eqnarray}}
\newcommand{\Ref}[1]{(\ref{#1})}
\newcommand{\tf}{\tilde{f}}
\newcommand{\tI}{\tilde{I}}
\newcommand{\hbv}{\widehat{{\bf v}}}
\newcommand{\hbk}{\widehat{{\bf k}}}
\newcommand{\bkplus}{{{\bf k}}_+}
\newcommand{\bkmin}{{{\bf k}}_-}
\newcommand{\bk}{{\bf k}}
\newcommand{\bq}{{\bf q}}
\newcommand{\bg}{{\bf g}}
\newcommand{\bgperp}{{{\bg}_\bot}}
\newcommand{\bn}{{\bf n}}
\newcommand{\bG}{{\bf G}}
\newcommand{\rg}{{g}}
\newcommand{\gperp}{{g_\bot}}
\newcommand{\gpar}{{{g}_\parallel}}
\newcommand{\hgpar}{{\hat{g}_\parallel}}
\newcommand{\bv}{{\bf v}}
\newcommand{\bc}{{\bf c}}
\newcommand{\bw}{{\bf w}}
\newcommand{\br}{{\bf r}}
\newcommand{\bu}{{\bf u}}
\newcommand{\half}{\textstyle{\frac{1}{2}}}
\newcommand{\eps}{{\epsilon}}
\newcommand{\NF}{\text{NF}}
\newcommand{\WN}{\text{WN}}
\newcommand{\bull}{{$\bullet$}}
\newcommand{\nn}{\nonumber \\}
\newcommand{\XXX}{{\bf XXXX}}

\title{
Quasi-elastic solutions to the nonlinear Boltzmann equation for
dissipative gases}

\author{A. Barrat}
\affiliation{Laboratoire de Physique Th\'eorique (UMR du CNRS 8627),
B\^atiment 210,
Universit\'e Paris-Sud, 91405 Orsay (France)}
\author{E. Trizac}
\affiliation{Universit\'e
Paris-Sud, LPTMS, UMR 8626, Orsay Cedex, F-91405 and
CNRS, Orsay, F-91405
}
\author{M.H. Ernst}
\affiliation{Theoretische Fysica, Universiteit Utrecht, Postbus 80.195,
3508 TD Utrecht (The Netherlands)}

\begin{abstract}
The solutions of the one-dimensional homogeneous nonlinear Boltzmann
equation are studied in the QE-limit (Quasi-Elastic; infinitesimal
dissipation) by a combination of analytical and numerical techniques.
Their behavior at large velocities differs qualitatively from that for
higher dimensional systems. In our generic model, a dissipative fluid is
maintained in a non-equilibrium steady state by a stochastic or
deterministic driving force. The velocity distribution for stochastic
driving is regular and for infinitesimal dissipation, 
has a stretched exponential
tail, with an unusual stretching exponent $b_{QE} = 2b$, twice as large
as the standard one for the corresponding $d$-dimensional system at finite
dissipation. For deterministic driving the behavior is more subtle and
displays singularities, such as multi-peaked velocity distribution functions. 
We classify the
corresponding velocity distributions according to the nature and scaling
behavior of such singularities.
\end{abstract}


\maketitle

\section{Introduction}

\subsection{Background and outline}
The model of inelastic hard spheres is one of the most simple
frameworks to describe granular gases (see e.g.
\cite{Gold,PoschelBrill,usJPCM} for reviews and further
references).  The contraction of phase space due to dissipative
collisions leads to a non-equilibrium behavior that is markedly
different from that of equilibrium systems (non-Gaussian velocity
distributions, counter-intuitive hydrodynamics, breakdown of
kinetic energy equipartition etc \cite{Gold,PoschelBrill,usJPCM}).
In this paper, we study in detail the limit of quasi-elasticity
\cite{Mcnamara2,Caglioti,Benedetto,Ramirez} with particular emphasis
on one-dimensional systems, that have already been the subject of
some interest \cite{Mcnamara1,Sela,bennaim,Rosas}.

The kinetic description is provided by the nonlinear Boltzmann equation.
As we are interested in the velocity statistics of dissipative gases, we
will restrict ourselves to homogeneous and isotropic solutions. Any
spatial dependence will therefore be discarded. The time evolution of the
velocity distribution function $F(v,t)$ is then governed by the following
equation\cite{EB-Rap02,ETB},
\be\label{eq:BE}
\partial_t F(v) + {\cal F}F
\,=\, I (v|F)\equiv \int_\bn \int d\bw \, g^\nu \left[
\frac{1}{\alpha_0^{\nu+1}} F(v^{**}) F(w^{**}) - F(v) F(w) \right].
\ee
Here ${\cal F}F$ represents the action of a driving mechanism, that
injects energy into the system, and counterbalances the energy dissipated
by inelastic collisions. Consequently, the system is expected to reach a
non-equilibrium steady-state. In the equation above ${\bg}=\bv-\bw$
denotes the relative velocity of colliding particles with $g=|\bg|$,
$\int_\bn (\cdots) = (1/\Omega_d) \int d\bn(\cdots)$ is an angular average
over the surface area \smash{$\Omega_d = 2 \pi^{d/2}/\Gamma(\half d)$} of
a $d$-dimensional unit sphere, and $g^\nu$ models the collision frequency.
Note that in one dimension the integral $\int_\bn$ is absent. We have
absorbed constant factors in the time scale. Here $(\bv^{**},\bw^{**})$
denote the restituting velocities that yield ($\bv,\bw$) as
post-collisional velocities, i.e.
\be
\bv^{**} \,=\,  \bv - \half(1+\alpha_0^{-1})(\bg\cdot\bn)\bn \quad; \quad
\bw^{**} = \bw +  \half(1+\alpha_0^{-1})(\bg\cdot\bn)\bn
\label{eq:colllaw},
\ee
where (unit) vector $\bn$ is parallel to the line of centers of the
colliding particles. Note that $\bn\bn$ is  replaced by $1$ in one
dimension. The direct collision law is obtained from (\ref{eq:colllaw}) by
interchanging pre- and post-collision velocities and by replacing
$\alpha_0 \to 1/\alpha_0$ where $\alpha_0<1$ is the restitution
coefficient. Each collision leads to an energy loss proportional to
$(1-\alpha_0^2)$. Elastic collisions therefore correspond to $\alpha_0=1$.

 In this article, we will consider the source term in (\ref{eq:BE})
to be of the form
\be \label{source} {\cal F} F = {\partial} \cdot
({\bf a}F) - D \partial^2 F = \gamma {\partial} \cdot (\hat{\bv} v^\theta
F) - D {\partial}^2 F,
\ee
where ${\bf a} = \hat{{\bf v}}v^\theta$ is a negative friction force,
$\boldmath{\partial} \equiv \partial /\partial \bv$ is the gradient in
$\bv$-space, and $\gamma$ and $D$ are positive constants. Two situations
will be addressed : ($\gamma=0$, $D>0$) or ($\gamma >0$, $D=0$). They
correspond respectively to {\it stochastic} White Noise (WN), or to {\it
deterministic} nonlinear Negative Friction (NF). While the WN driving
mechanism has been extensively studied \cite{WN,Benedetto,MS}, the
Negative Friction has been introduced more recently \cite{ETB}. The
continuous exponent $\theta  \geq 0$ selectively controls the energy
injection mechanism. Schematically, increasing the value of $\theta$
corresponds to injecting  more energy  in the large velocity tail of the
distribution. However two special values have been studied in the past
\cite{MS,BBRTvW,EB-Rap02,SE-PRE03,ETB-EPL06,ETB}, i.e. 
(i) the Gaussian thermostat ($\theta=1$), which is equivalent
to the homogeneous free cooling state, where the system is unforced and
the possibility of spatial heterogeneities discarded (see e.g. \cite{MS}),
and (ii) the case $\theta=0$, referred to as 'gravity thermostat'
\cite{MS}, or as 'negative solid friction' \cite{ETB-EPL06,ETB} .

We emphasize that $\gamma$ and $D$ are irrelevant constants that can be
eliminated (see below), whereas  the exponents $\theta$ and $\nu$ are
fundamental quantities for our purposes. As it appears in Eq.
(\ref{eq:BE}), $\nu$ governs the collision frequency of the system:
$\nu=1$ corresponds to hard-sphere like dynamics and $\nu=0$ to the
so-called Maxwell model \cite{ME,Maxwell,Balda,EBri,EB-Rap02}. Our
collision kernel generalizes these two  cases to a general class of
repulsive power law potentials, where $\nu$ is related to the power law
exponent and the dimensionality (see  \cite{ETB}).

Under the action of the driving term $\cal F$, the solution of
(\ref{eq:BE}) evolves towards a non-equilibrium steady state. We will be
interested in the properties of the corresponding velocity distribution
$F(v)$, in the limit where $\alpha_0 \to 1^-$. Since the limit $\alpha_0
\to 1$ turns out to be singular in one dimension, attention must be paid
to the fact that the value $\alpha_0=1$ (elastic interactions) has to be
analyzed separately. Indeed, when $\alpha_0=1$ in one dimension, the
collision law (\ref{eq:colllaw}) simply corresponds to an exchange of
particle labels. So the initial velocity distribution does not evolve in
time. On the other hand, the steady state velocity distribution at any
$\alpha_0<1$ is independent of the initial condition. In dimensions higher
than 1, this property holds for all values of $\alpha_0\leq 1$.  In other
words, the one dimensional situation with $\alpha_0<1$ exhibits universal
features, unlike its elastic counterpart. The quasi-elastic limit
$\alpha_0\to 1^-$ is therefore peculiar since a point with no universal
properties ($\alpha_0=1$) is approached via a `universal' route
($\alpha_0<1$). The resulting behavior of $F$ shows some surprising
features, that may be considered as mathematical curiosities, but are
analytically challenging. It also turns out that they are numerically
difficult to study.  The numerical study relies on the DSMC (Direct
Simulation by Monte Carlo) algorithm \cite{Bird}, which allows us to
obtain an exact numerical solution of the Boltzmann equation. As $\alpha_0$
approaches $1$, the memory of the initial conditions lasts for longer and
longer times so that the computer time needed to reach the non-equilibrium
steady state increases and simulations become more and more
time-consuming.

The behavior of our system is somewhat simpler in the case where energy is
injected by a stochastic force (WN), and we start by analyzing this
driving mechanism in section \ref{sec:WN}. It will be shown that the
regular high energy tail of $F(v) \sim \exp[-v^{b}]$, which holds in any
space dimension \cite{EB-Rap02,ETB-EPL06,ETB}, is preempted by a
quasi-elastic tail characterized by a different stretching exponent
$b_{QE} =2 b$. This is a signature of the non-commutativity of the
limits: $v\to \infty$ and $\alpha_0\to 1^-$ . The case of driving through
negative friction will be addressed in detail in section \ref{sec:NF}. As
already observed for the homogeneous cooling state of inelastic hard rods
\cite{Mcnamara2,Caglioti,Sela,BBRTvW,SE-PRE03}, the velocity distribution
becomes singular in the QE-limit, where it may approach a multi-peaked
solution, and not a Gaussian. Starting from a small-inelasticity expansion
for the collision operator $I(v|F)$ in (\ref{eq:BE}), we characterize the
scaling behavior.  By a combination of analytical work and numerical
evidence, we propose a classification of the different types of limiting
velocity distributions, several of which correspond to new types of
solutions of the nonlinear Boltzmann equation.

\subsection{Preliminary remarks}

We start by introducing some notations and summarizing a few results
\cite{ETB} that are relevant for our study. In the subsequent analysis, it
is convenient to introduce the variables $p=(1+\alpha_0)/2$,
$q=(1-\alpha_0)/2$ so that $p+q=1$, and to measure the velocities in units
of the r.m.s. velocity. We study steady states and introduce a rescaled
velocity distribution $f(c)$ such that $F(v) = v_0^{-d} f(v/v_0)$ where
$v_0$ is the r.m.s velocity, $c=v/v_0$, and $d$ is the number of spatial
dimensions. By definition, $\int d\bc f=1$ and the normalization chosen
reads $\int d\bc c^2 f=d/2$. After inserting the scaling ansatz in
Eq.(\ref{eq:BE}), we have shown in \cite{ETB} that a stable steady state
for WN driving is reached provided $b_{WN}=1+\nu/2>0$, and for NF driving
provided $b_{NF} = \nu + 1- \theta >0$. When $b<0$, the non-equilibrium
steady state is unstable, i.e. it is a repelling fixed point of the
dynamics. Our analysis should therefore be limited to the cases $\nu>-2$
for WN and to $\nu>\theta-1$ for NF.

We have shown in \cite{ETB} that the quantity $b$ introduced above not
only separates stable from unstable situations, but also governs the high
energy tail of $f(c)$. In marginal cases where $b$ vanishes, $f$ has a
power-law tail. The freely cooling ($\theta=1$) Maxwell model ($\nu=0$)
provides an illustration that has been discussed in
\cite{Maxwell,Balda,EBri}. On the other hand, when $b>0$, $f$ has a
stretched exponential tail so that $\ln f(c) \propto -c^b$ at large $c$.
This result holds in any dimension. In $d=1$, the corresponding tail may
however be `masked' when $\alpha_0$ is close to unity, a phenomenon
already observed for hard rods ($\nu=1$) in \cite{BBRTvW}: the behavior
$\ln f(c) \propto -c^b $ holds for $c>c^*(\alpha_0)$, where
$c^*(\alpha_0)$ is an $\alpha_0$-dependent threshold. In the QE-limit, the
threshold value $c^*(\alpha_0) \to \infty$. As a consequence, considering
the limit of large $c$ at any finite $\alpha_0$, the standard behavior
with exponent $b$ is observed. Alternatively, taking first the limit
$\alpha_0\to 1^-$ at fixed $c$ , {\em new} tails may appear. It is the
purpose of the present paper to study their properties.  In addition, in
the case of NF driving, $f(c)$ becomes singular in the quasi-elastic
limit. Our goal will then be to understand the underlying scaling behavior
and to propose a classification of the various limiting shapes for the
velocity distribution.

\section{White Noise (WN) driving}
\label{sec:WN}

Unless explicitly stated, we limit ourselves to $d=1$. 
In the case of WN driving the 
integral equation (\ref{eq:BE}) for the scaling form, $f(c)
=v_0 F(c v_0)$, becomes
with the help of the relation $I(v|F)=v_0^{\nu-1} I(c|f)$:
\be \label{BE-WN}
I(c|f) =-Dv_0^{-\nu-1}f^{\prime\prime}(c)= -\half pq \kappa_\nu
f^{\prime\prime}(c).
\ee
The second equality has been obtained by applying $\int dc c^2 (\cdots)$
to the first equality, see \cite{ETB}), yielding
\be \label{D-WN}
Dv_0^{-\nu-1}= -\half \langle\langle c^2I(c|f)\rangle\rangle =\half pq
\langle\langle |c-c_1|^{\nu+2}\rangle\rangle \equiv \half pq \kappa_\nu,
\ee
where the double brackets denote an average with weight $f(c) f(c_1)$.
Similarly, simple brackets denote an average with weight $f(c)$. 

It is next convenient to treat the 1-D collision term,
\be I(c|f)\label{NL-coll}
= \int d c_1 |c-c_1|^\nu \left[ {\alpha_0^{-\nu-1}} f(c^{**}) f(c_1^{**}) -
f(c)f(c_1) \right] \ ,
\ee
by using the inverse transformation, $c=qu + pc_1^{**}$ and $c_1 =pu +q
c_1^{**}$,  where $u=c^{**}$ and $|c-c_1|=\alpha_0|c-u|/p$. We obtain
\be \label{NL-coll1}
I(c|f) = \int du |c-u|^\nu
\left[ p^{-\nu-1} f(u)f\left( \frac{c-qu}{p} \right) -
f(u)f(c)\right].
\ee
In the quasi-elastic limit $q \to 0^+$, $p \to 1^-$, we perform the
small-$q$ expansion,  following Refs.\cite{Mcnamara2,Caglioti,Benedetto},
\be \label{help-WN}
\frac{1}{p^{\nu+1}}f\left( \frac{c-qu}{p} \right) = (1+ (\nu+1)q)f(c) +q
(c-u)f'(c)  +{\cal O}(q^2).
\ee
Then we find to ${\cal O} (q)$ included,
\ba \label{QE-coll}
 I(c|f)&=& q \int du f(u) |c-u|^\nu \left[ (c-u)f'(c) + (\nu+1)f(c) \right]
\nn &=&  q  \frac{d}{dc} f(c)\int du |c-u|^\nu (c-u)f(u) .
\ea
Note that these results hold irrespective of the driving mechanism. The second
equality can be verified by evaluating the derivative. 
Inserting  of (\ref{QE-coll}) into 
(\ref{BE-WN}) allows us to integrate \Ref{BE-WN} once, yielding:
 \be
 \label{QE-WN-eq}
 f'(c) +  f(c)(2/\kappa_\nu) \int d u |c-u|^{\nu}(c-u) f(u) =0 .
\ee
This equation can be integrated once more to obtain the implicit equation,
\ba \label{f-asympt}
f(c) &=& {\cal C}\exp \left( - \frac{2}{(\nu+2)\kappa_{\nu} } \int
du |c-u|^{\nu+2} f(u) \right)
\nn &\sim & {\cal C} \exp\left( -\frac{2c^{\nu+2}}{(\nu+2)\kappa_{\nu} }\right)
\qquad (  c \to \infty),
\ea
where ${\cal C}$ is an integration constant. The large velocity tail
immediately follows.  We recover a result already obtained in
\cite{Benedetto,BBRTvW} for hard rods ($\nu=1$), as well as one for
Maxwell models $(\nu=0)$ in \cite{SE-PRE03}, and we note that in the
QE-limit the exponent $b_{QE} = 2b_{WN} = \nu +2$. On the other hand, for
any finite $\varepsilon=1-\alpha_0$, the large $c$-behavior is given by $\ln
f(c)\propto -c^b$ with  $b=1+\nu/2$ \cite{ETB}.
These two facts show that the limits $c \to \infty$ and $\varepsilon \to 0$
do not commute,
\begin{eqnarray} \label{St+QE-tail}
f(\varepsilon,c) \,\,\sim & \,\, \exp(-A \,c^{1+\nu/2}) \quad
&\hbox{(standard tail: large $c$, fixed $\varepsilon \neq 0$)}\nn
 f(\varepsilon, c) \sim& \,\, \exp(-A\, c^{\nu+2}), \quad
 &\mbox{(QE-tail: small $\varepsilon$, fixed $c$)}.
\end{eqnarray}
or, in a more rigorous formulation
\begin{eqnarray}
\lim_{\smash{\varepsilon \to 0^+}}  ~\lim_{c\to \infty} \frac{\ln
f(\varepsilon,c)}{c^{1+\nu/2}} &~=~ {\cal C} \qquad &(\mbox{standard tail})
\nn \lim_{c\to \infty} ~ \lim_{\smash{\varepsilon\to  0^+}} \frac{\ln
f(\varepsilon,c)}{c^{\nu+2}}& ~= ~ {\cal C}^\prime \qquad &(\mbox{QE-tail}).
\end{eqnarray}.

\begin{figure}[htb]
\vskip .5cm
\includegraphics[height=6cm,angle=0]{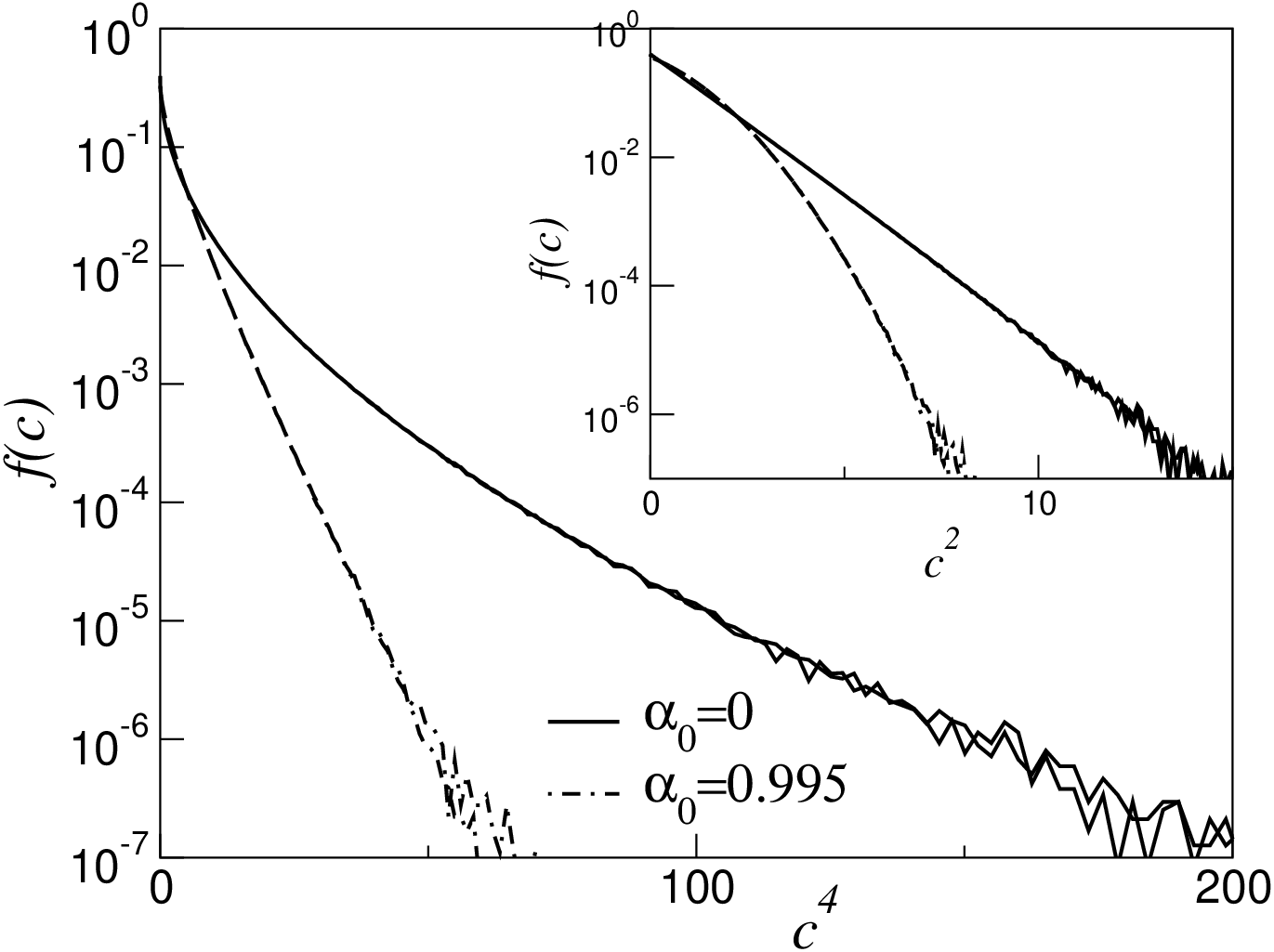}
\caption{
Rescaled velocity distribution $f(c)$ obtained by solving numerically  the
1-D Boltzmann equation (\ref{eq:BE}) with WN driving by the DSMC technique
\cite{Bird} for the strongly interacting so-called 'very hard particles'
model
($\nu = 2$) \cite{ME}). Two extreme cases have been simulated, one close
to  perfect elasticity ($\alpha_0=0.995$) where the QE-tail $\ln f$ vs.
$c^{\nu+2}$ (main frame) is visible, and the other at complete
inelasticity ($\alpha_0=0$), where the standard tail  $\ln f$ vs.
$c^{1+\nu/2}$ (inset) is visible.} \label{fig:WN1}
\end{figure}

\begin{figure}[htb]
\vskip .5cm
\includegraphics[height=5cm,angle=0]{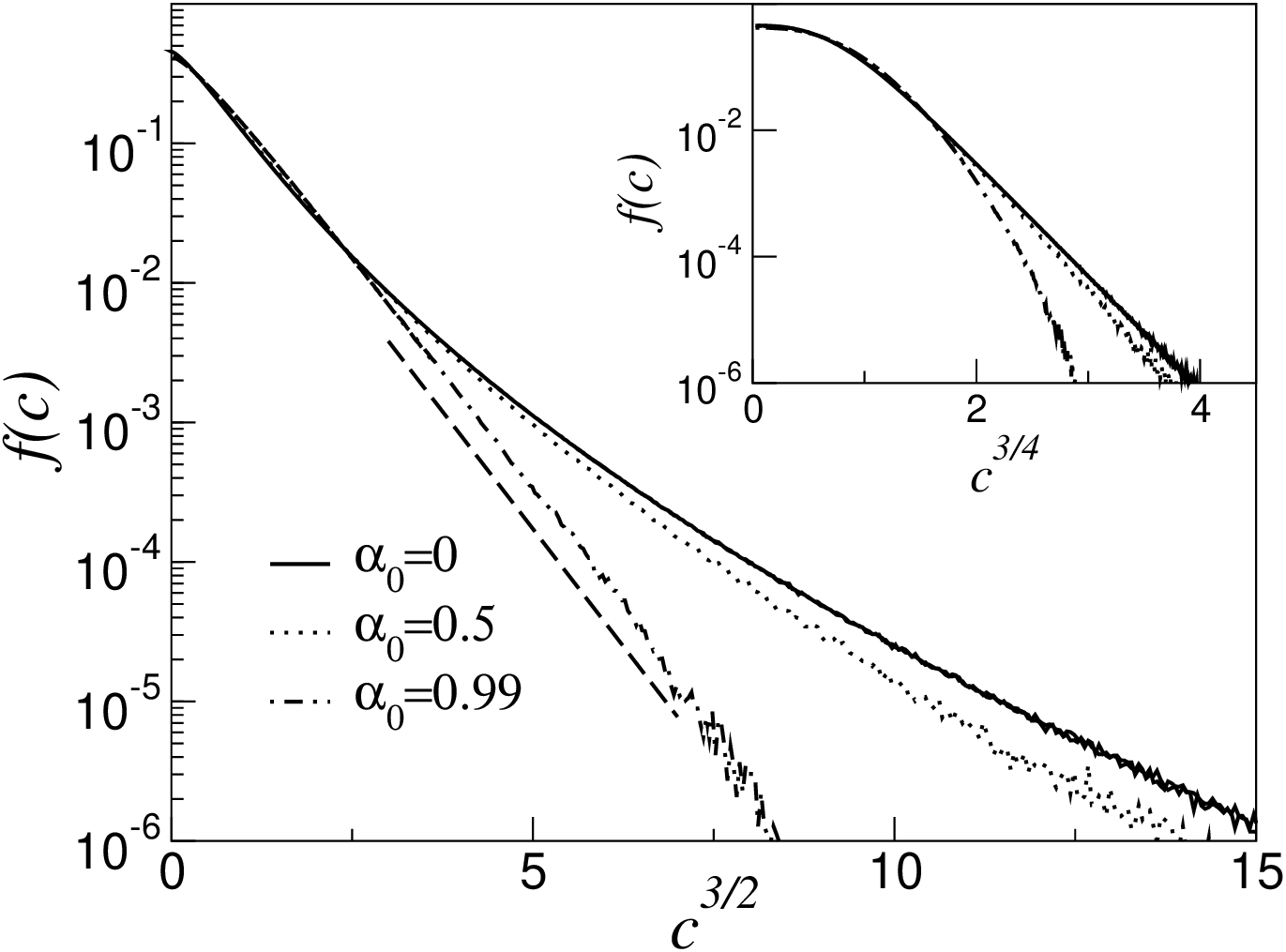}
\hskip .5cm
\includegraphics[height=5cm,angle=0]{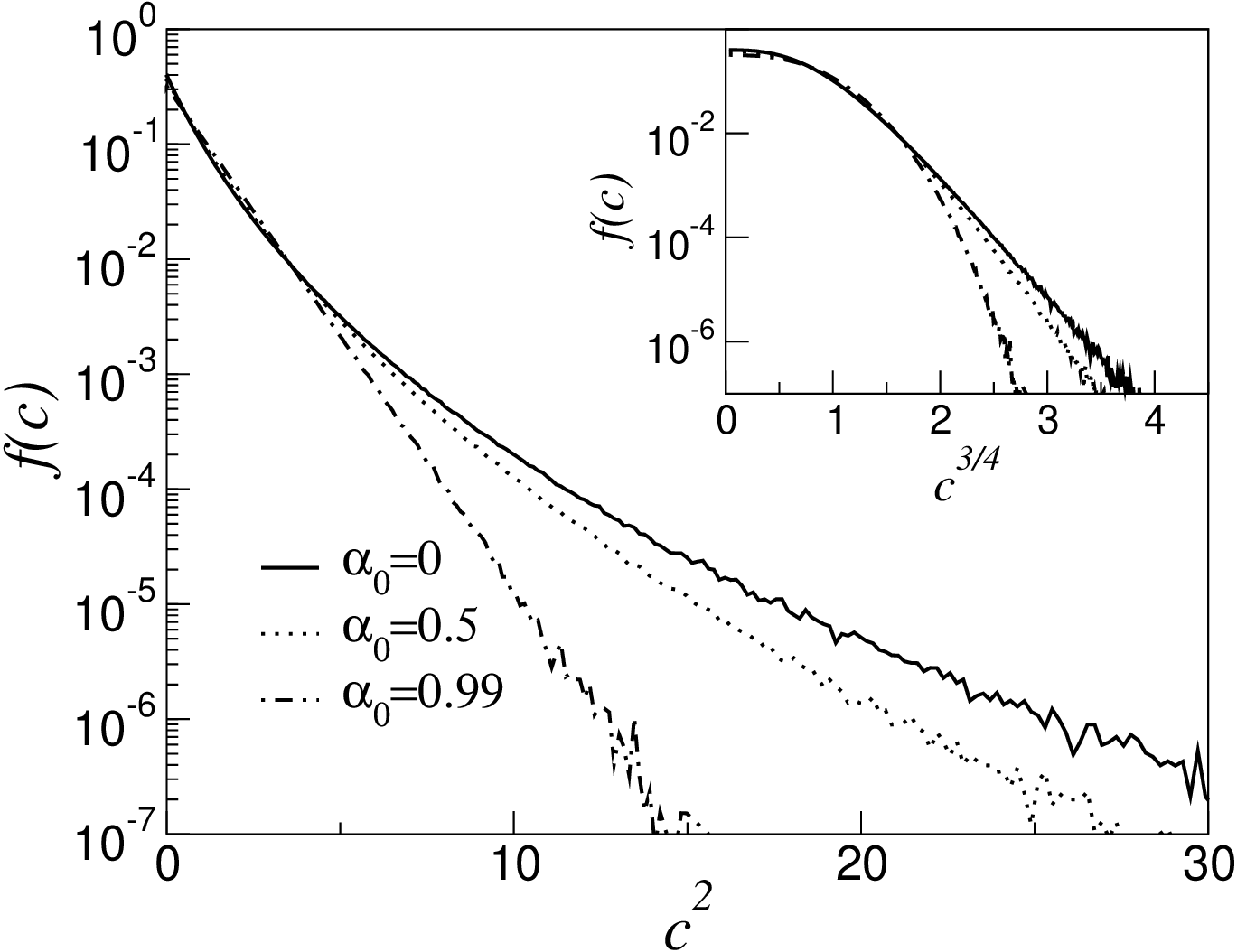}
\caption{Left: Velocity distribution obtained from DSMC simulations for weakly
interacting 1-D particles $(\nu=-\half)$ with $\ln f(c)$ vs. $c^{\nu+2}$
(main frame) and vs. $c^{1+\nu/2}$ (inset). In the main frame the dashed
line is a guide for the eye corresponding to the expected behavior
(\ref{St+QE-tail}). Right: $d=2$, $\ln f(c)$ vs $c^{2}$ (main) and vs.
$c^{1+\nu/2}$ (inset). } \label{fig:WN2}
\end{figure}

We have successfully tested these predictions against Direct Simulation
Monte Carlo  data \cite{Bird}. Figure \ref{fig:WN1} shows  that the
standard tail with exponent $ b_{QE}= 1+\nu/2$ is clearly observed for
$\alpha_0=0$ (see inset), while the QE-tail with exponent $\nu+2$ applies
for $\alpha_0=0.995$ (see main frame). 
Figure \ref{fig:WN2} conveys a similar message, and shows
further that in two dimensions, Gaussian behavior is recovered, as
expected, when $\alpha_0\to 1$ (see the inset of the right hand side
figure). This illustrates the qualitatively different nature of the
quasi-elastic limit in $d=1$, as opposed to higher dimensions.

\section{Negative Friction (NF) driving}
\label{sec:NF}

\begin{figure}[htb]
\vskip .5cm
\includegraphics[height=5cm,angle=0]{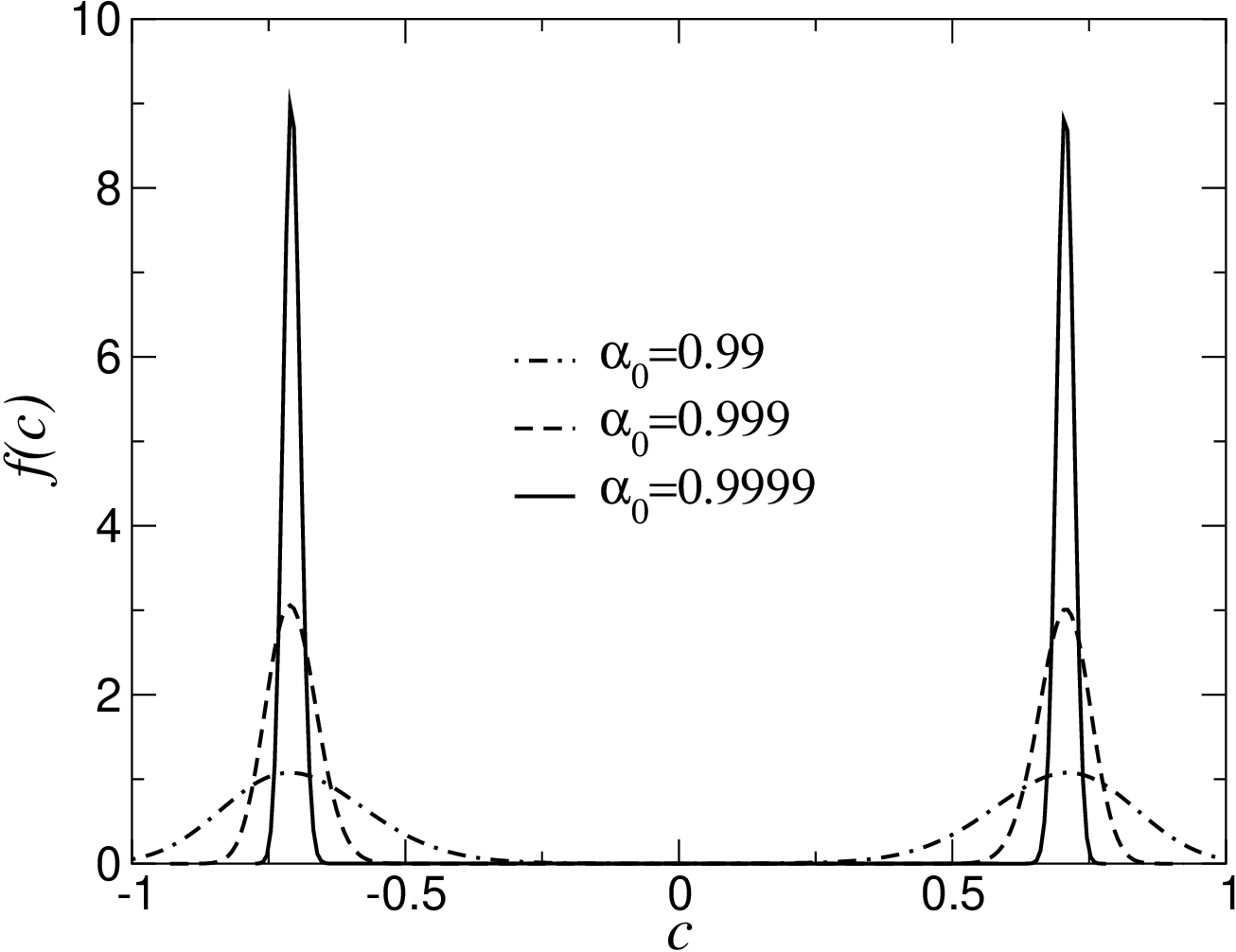}
\caption{Velocity
distribution obtained from DSMC simulations with NF driving for
$(\theta,\nu) = (1.0, 1.2)$ in 1-D. As $\varepsilon= 1-\alpha_0 \to 0^+$,
$f(c)$ becomes increasingly peaked around  $c =\pm c^* \equiv \pm
1/\sqrt{2}$, and ultimately evolves into two symmetric Dirac
distributions. All distributions displayed here and in other figures are
such that $\int dc f=1$, $\int dc c f=0$ and $\int dc c^2 f=1/2$. }
\label{fig:2peaks}
\end{figure}

\begin{figure}[t]
\vskip .5cm
\includegraphics[height=5cm,angle=0]{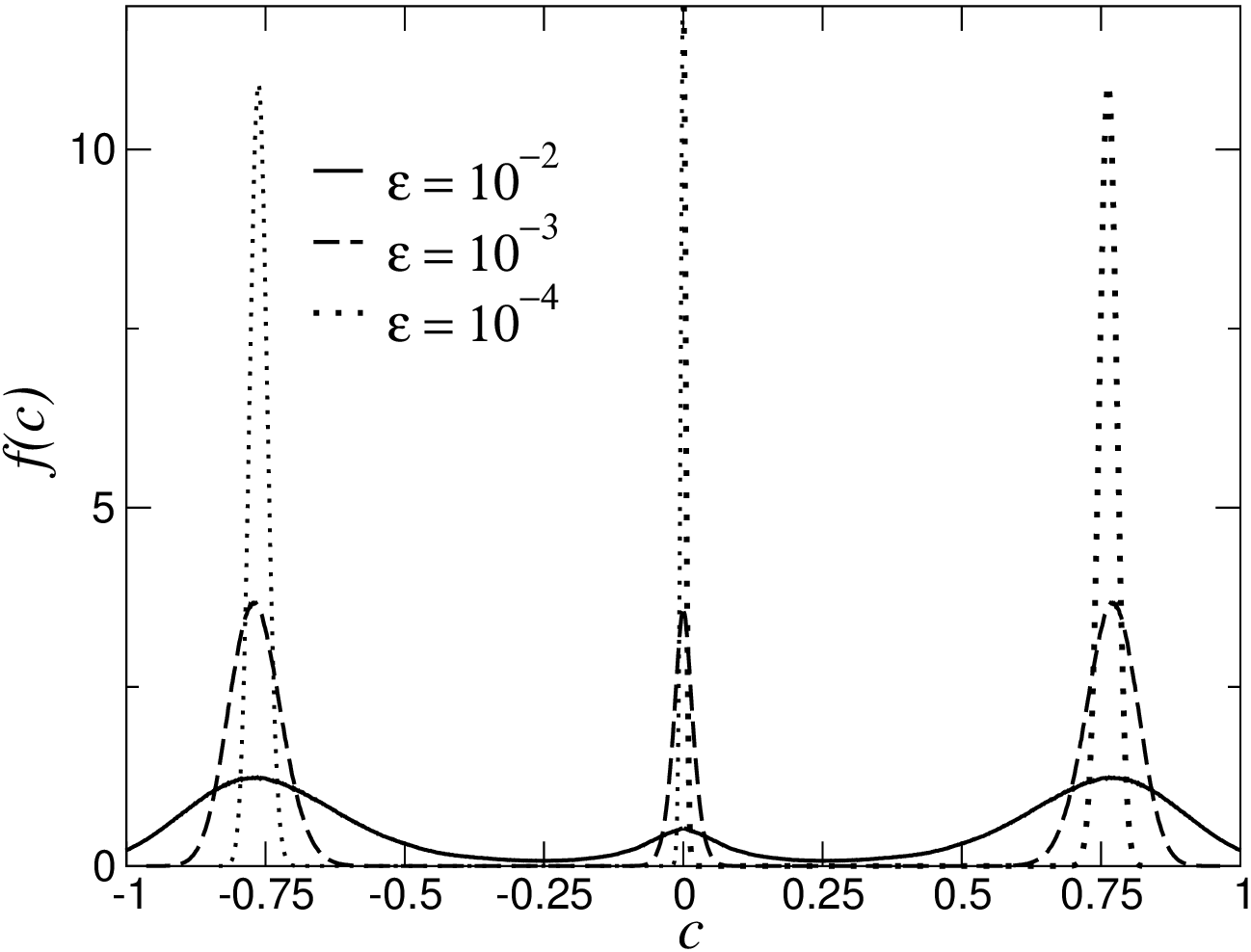}
\vskip -4.5mm \caption{ Same as Figure \ref{fig:2peaks} for slightly different
parameters ($\theta,\nu) = (1.1, 1.3$). The velocity distribution now
reaches a three-peak structure when $\varepsilon\equiv1-\alpha_0 \to 0^+$.}
\label{fig:3peaks}
\end{figure}

The scaling equation for the analog of \Ref{BE-WN} with NF driving
becomes,
\be \label{BE-NF}
I(c|f) = \frac{\gamma }{v_0^{\nu+1-\theta} } \frac{d}{dc} \left( c
|c|^{\theta-1} f \right) = \frac{pq \kappa_{\nu}}{2 \mu_{\theta+1}}
\frac{d}{dc} \left( c |c|^{\theta-1} f \right).
\ee
Here $I(c|f)$ takes the form (\ref{QE-coll}), and the analog of \Ref{D-WN}
has been used to eliminate $\gamma$. Moreover, $\mu_{\theta+1}=\langle
|c|^{\theta+1} \rangle$ and $\kappa_{\nu}= \langle\langle |c-c_1|^{\nu+2}
\rangle\rangle$. We therefore have to ${\cal O}(q)$ included,
\be \label{QE-NF}
f(c) \int d c_1 |c-c_1|^{\nu}(c-c_1) f(c_1) = \frac{\kappa_{\nu}}{2
\mu_{\theta+1}} c |c|^{\theta-1} f(c).
\ee
Previous studies of the free cooling regime of hard rods ($\theta=1, \,
\nu=1$) have shown that $f(c)$ becomes singular when $\alpha_0\to 1^-$,
and evolves into two symmetric Dirac peaks
\cite{Mcnamara2,Sela,Caglioti,BBRTvW}.  It is indeed easy to check that
$f(c)=[\delta(c+a)+\delta(c-a)]/2$ is a solution of \Ref{QE-NF}), with
$a=1/\sqrt{2}$ as required by our choice of normalization, $\langle
c^2\rangle=1/2$, and $\kappa_{\nu}/(2 \mu_{\theta+1})=2^\nu
a^{\nu+1-\theta}$.  Figure \ref{fig:2peaks} shows that the approach to
such a solution may be observed in the numerical simulations for
$(\theta,\nu)\neq (1,1)$.  In addition, one can check that $f(c)=A
[\delta(c+a)+\delta(c-a)] +B \delta(c)$ is also a solution of Eq.
(\ref{QE-NF}), provided that $2A+B=1$ and $4Aa^2=1$ to enforce
normalization. The DSMC results may indeed display such a three-peak
structure (see Figure \ref{fig:3peaks}). Note that the parameters
corresponding to  Figures \ref{fig:2peaks} and \ref{fig:3peaks} are quite
close: $(\theta,\nu) = (1.0, 1.2)$ and (1.1, 1.3) respectively.

\begin{figure}[t]
\vskip .5cm
\includegraphics[height=5cm,angle=0]{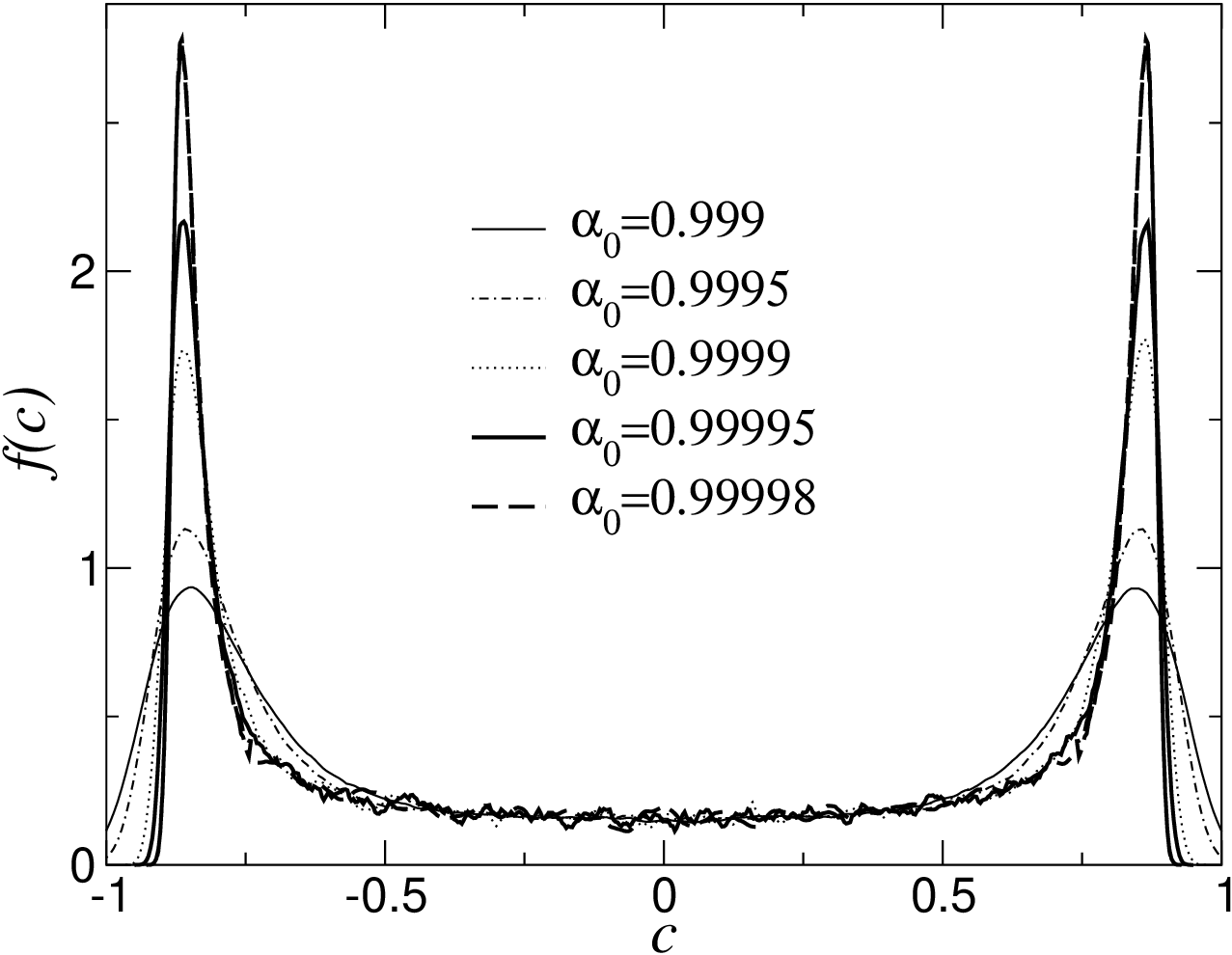}
\vskip -4.5mm \caption{
Velocity distribution obtained from DSMC for
$(\theta,\nu )= (1.0, 0.5)$. These parameters correspond to the border of
the ``Zoo'' region shown in Figure \ref{fig:diagram}.} \label{fig:complex}
\end{figure}

However, upon changing the parameters $\theta$ and $\nu$, it appears that
more complex shapes can be observed: $f(c)$ may evolve towards a 4-, 5-,
6-peaks form, or other structures such as displayed in Figure
\ref{fig:complex} where $f(c)$ seems to diverge at some points when
$\alpha_0\to 1^-$, with nevertheless a finite support. The diversity of
the various velocity distributions obtained numerically calls for a
rationalizing study.  By a combination of analytical work and numerical
evidence, we will propose below a classification of the different possible
limiting velocity distributions. In addition, in the double peak case, two
natural questions will be addressed: Are the peaks exemplified in Figure
\ref{fig:2peaks} self-similar~? If so, what is their shape~?

\subsection{The double peak scenario : structure and scaling}
\label{sec:scal2peaks}

\begin{figure}[t]
\vskip .5cm
\includegraphics[height=5cm,angle=0]{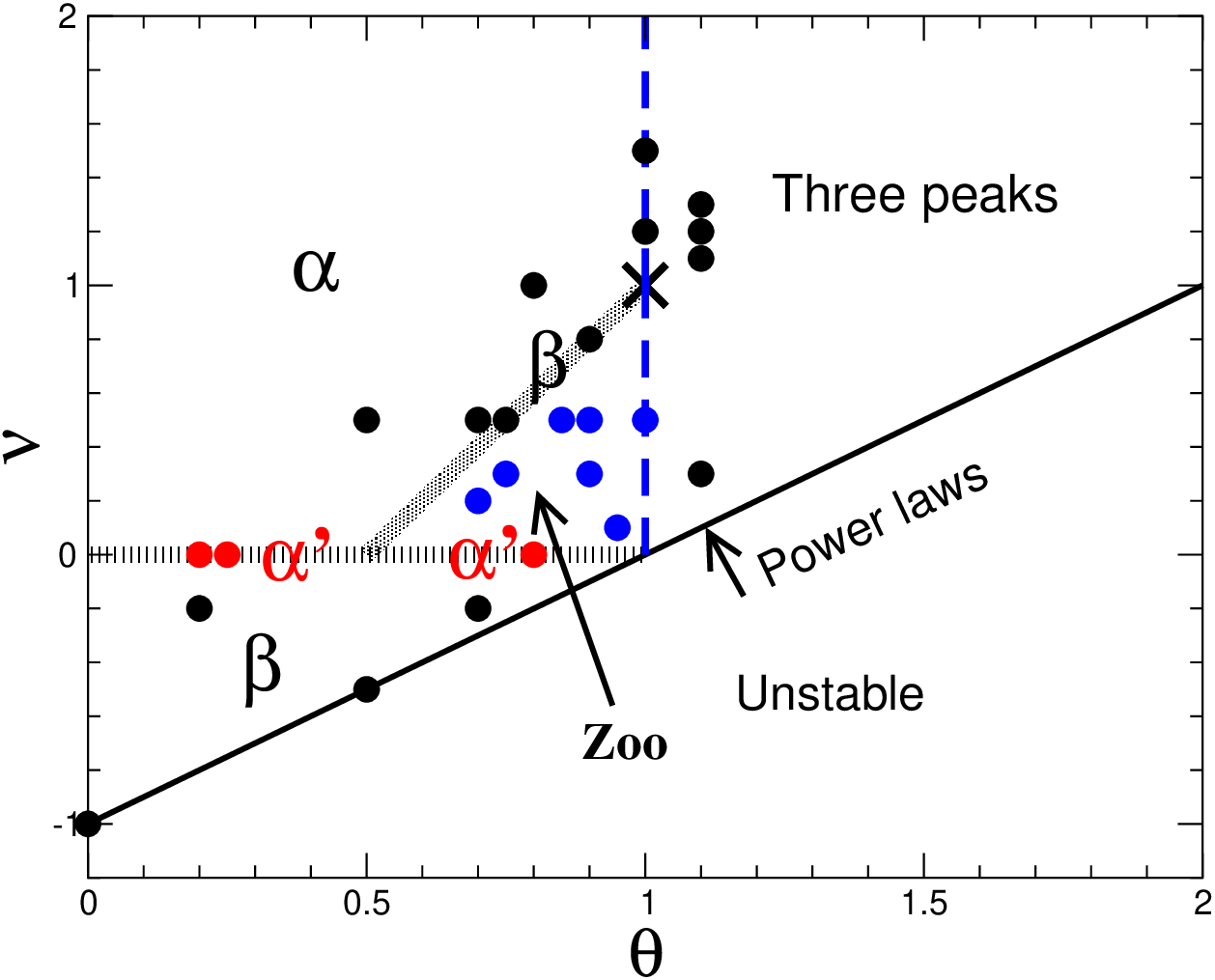}
\vskip -4.5mm \caption{Phase diagram in the quasi-elastic limit. Each dot
corresponds to a set of numerical simulations at smaller and smaller
values of $\varepsilon=1-\alpha_0$, performed to check the validity of the
scaling behaviors summarized in Table \ref{table:table}.  The stability of
the non-equilibrium steady state requires that $\nu>\theta-1$, while on
the diagonal $\nu=\theta-1$ the velocity distribution has a power-law tail
\cite{ETB}. Here $\alpha$-, $\alpha'$- and $\beta$-scalings are associated
with a distribution with two Dirac peaks. For $\theta>1$ we observed
numerically a solution with three Dirac peaks, while there does not seem
to be a simple common feature for the distributions in the triangular
``Zoo'' region.  An example of type $\alpha$-scaling is given in Figure
\ref{fig:alpha} (See also Figure \ref{fig:beta1} for type $\beta$, and
Figure \ref{fig:alphaprime} for type $\alpha'$).  Figures \ref{fig:Zoo1}
and \ref{fig:Zoo2} together with Figure \ref{fig:complex} above give an
overview of several scaling shapes encountered in the Zoo region. The
cross at (1,1) corresponds to the homogeneous cooling state of inelastic
hard rods, explored in \cite{BBRTvW}. } \label{fig:diagram}
\end{figure}

We start by looking for scaling solutions to Eq. (\ref{eq:BE}), and
restrict our analysis to the limiting form with the symmetric double peak.
As in \cite{BBRTvW} we take $f(c)$ of the doubly peaked form,
\be \label{eq:ansatz}
f(c) = \frac{b}{4E} \left[ \psi \left( \frac{1+bc}{2E} \right)+ \psi
\left( \frac{1-bc}{2E} \right) \right],
\ee
where the width $E=q^\omega$ is expected to vanish when $q\to 0^+$ and
$b$, $\omega$, together with the function 
$\psi$ are unknowns. We impose $\langle 1
\rangle_\psi\equiv \int \psi=1$, and we choose $\langle x \rangle_\psi=0$,
which together with the condition $\langle c^2 \rangle=1/2$,  implies $b^2
= 2(1+4E^2 \langle x^2 \rangle_\psi)$, where normalization requires that
$b\to ({\sqrt 2})^- $ as $q\to 0^+$. We note that asymmetric forms with 
$\psi(x) \neq \psi(-x)$ may be realized (see \cite{BBRTvW} and later sections).
The
ansatz (\ref{eq:ansatz}) allows us to resolve the structure of the Dirac
peaks, shown in Figure \ref{fig:2peaks}, and to identify the type of
self-similar behavior involved. However to this end, we need to expand the
collision operator $I(c|f)$ to second order in the inelasticity $ \varepsilon
= 1-\alpha_0  = 2q$. Restriction to first order, as done in
(\ref{QE-coll}), enables us to show that the double Dirac form is a
solution of the Boltzmann equation, but does not allow us to impose
constraints on the shape  of $\psi$ and on the scaling exponent $\omega$.
Technical details can be found in the appendix, where it is shown that the
equation fulfilled by $\psi(x)$ reads,
\ba
\nonumber q \psi'(x) + 2E^2(\nu+1-2\theta)x\psi(x)
+2E^{\nu+2}\psi(x)\langle |x-y|^\nu (x-y)\rangle_\psi \nn
 -E^3\nu(\nu+1)\psi(x)(x^2+\langle y^2\rangle_\psi) +2E^3 \left[
(\nu+1)(\nu+2)-2\theta(\theta+1)   \right] \langle y^2 \rangle_\psi \,
\psi(x)\nn +4E^3\theta(\theta-1) x^2 \psi(x) +2E^{\nu+3}\psi(x) \langle
\langle |y-z|^{\nu+2} \rangle \rangle +{\cal O}(Eq\psi)=0, \label{eq:psi}
\ea
where $y,z$ are dummy variables, and terms of ${\cal O}(Eq\psi) $  have
been neglected. This relation involves terms of various orders in
inelasticity. Given that $E=q^\omega$, these terms are ${\cal O}(q^a)$
with $a=1, 2\omega, (2+\nu)\omega, 3\omega, (3+\nu)\omega, \omega+1$.
Depending on the values of the parameters $\theta$ and $\nu$ one has to
distinguish various possibilities to characterize the phase diagram, i.e.
the physically allowed region of the $(\theta,\nu)$-parameter space. The
stability criteria  for the steady state (see \cite{ETB}) constrain the
phase diagram in Figure \ref{fig:diagram} to be inside the region [$\theta
\geq 0, \nu \geq \theta-1$].

\subsubsection{Case $\nu > 0$ ($\alpha$- and $\beta$-scalings).}

\begin{figure}[t]
\vskip .5cm
\includegraphics[height=5cm,angle=0]{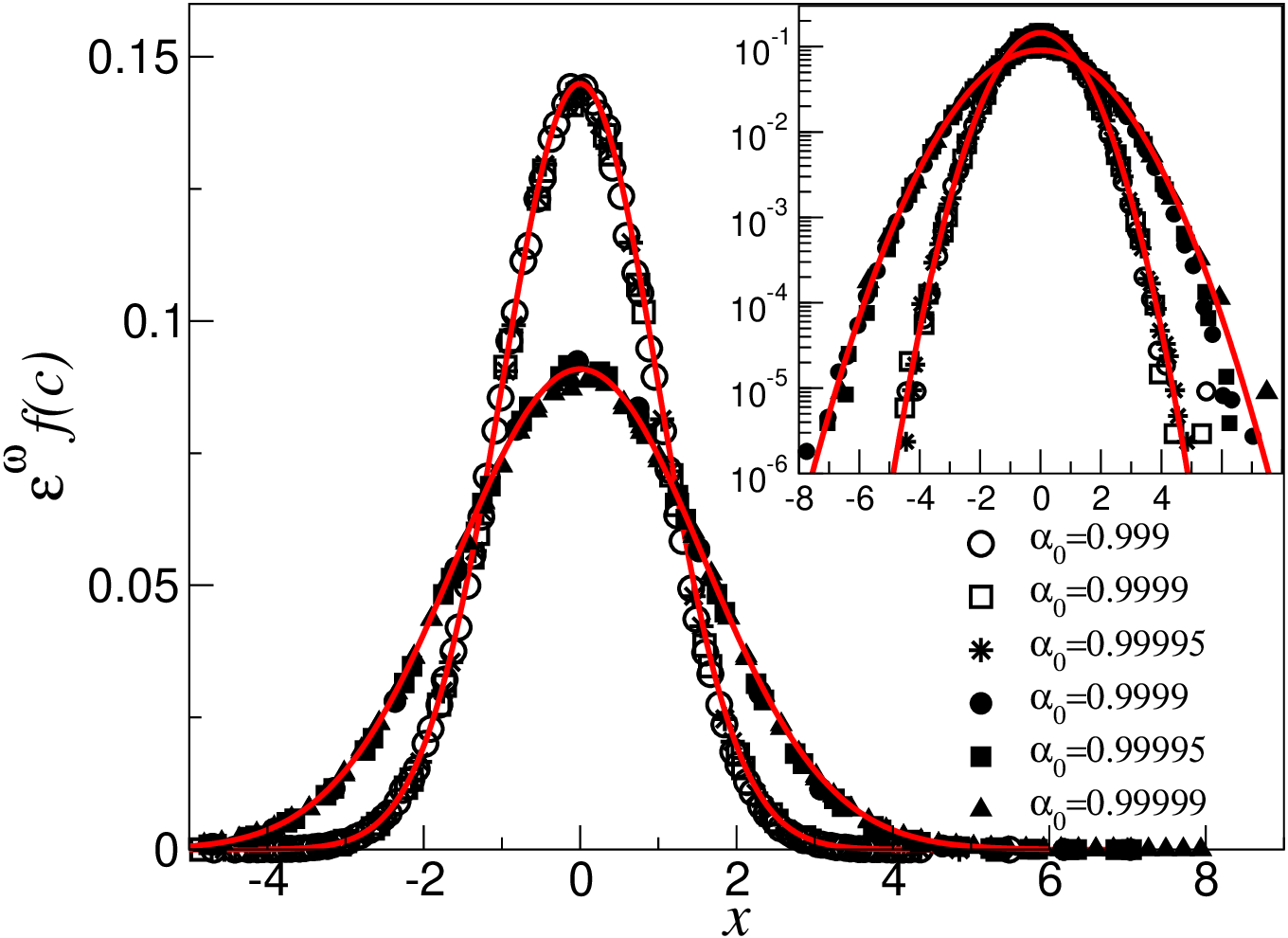}
\vskip -4.5mm \caption{Scaling
behavior of type $\alpha$. Plots of $\varepsilon^{\omega} f(c)$ vs. $x=
(|c|-c^*)/\varepsilon^{\omega}$ at $\theta=1$ for various inelasticities
$\varepsilon =1-\alpha_0$ with $\omega=1/2$ as predicted. Here $c^* =
1/\sqrt{2}$ corresponds to the peak of the distribution. Open symbols
correspond to $\nu=1.5$ and filled symbols are for $\nu=1.2$ (same as in
Figure \ref{fig:2peaks}). The inset shows the same results on a linear-log
scale. In each case, the prediction of Eq. (\ref{eq:psialpha}) for the
scaling function is shown by the continuous curve. Note that no fitting
parameter is involved.} \label{fig:alpha}
\end{figure}

{\it Type $\alpha$-scaling:} The terms of order $q$ and $q^{2\omega}$ are
the dominant ones in Eq. (\ref{eq:psi}). So, $\omega=1/2$ and
\be
\psi'(x) =-(\nu+1-2\theta)x\psi(x).
\ee
The scaling function $\psi$ is therefore Gaussian,
\be
\psi(x) \propto \exp \left( -(\nu+1-2\theta) x^2 \right).
\label{eq:psialpha}
\ee
Such a solution is meaningful only if $\nu+1-2\theta>0$. This scaling
behavior, hereafter referred to  as type $\alpha$ (not to be confused with
the restitution coefficient $ \alpha_0$) is compared in  Figure
\ref{fig:alpha} with Monte Carlo results. The agreement between the
analytical prediction and the numerical data involves two aspects : first,
the exponent $\omega=1/2$ allows us to rescale all distributions onto a
single master curve. Second, this curve is exactly of the form
(\ref{eq:psialpha}), where the prefactor hidden in the proportionality
sign $\propto$ follows from normalization. Note that the excellent
agreement between numerical data and analytical prediction is therefore
obtained without any fitting parameter.
\\ \\

\begin{figure}[t]
\vskip .5cm
\includegraphics[height=5cm,angle=0]{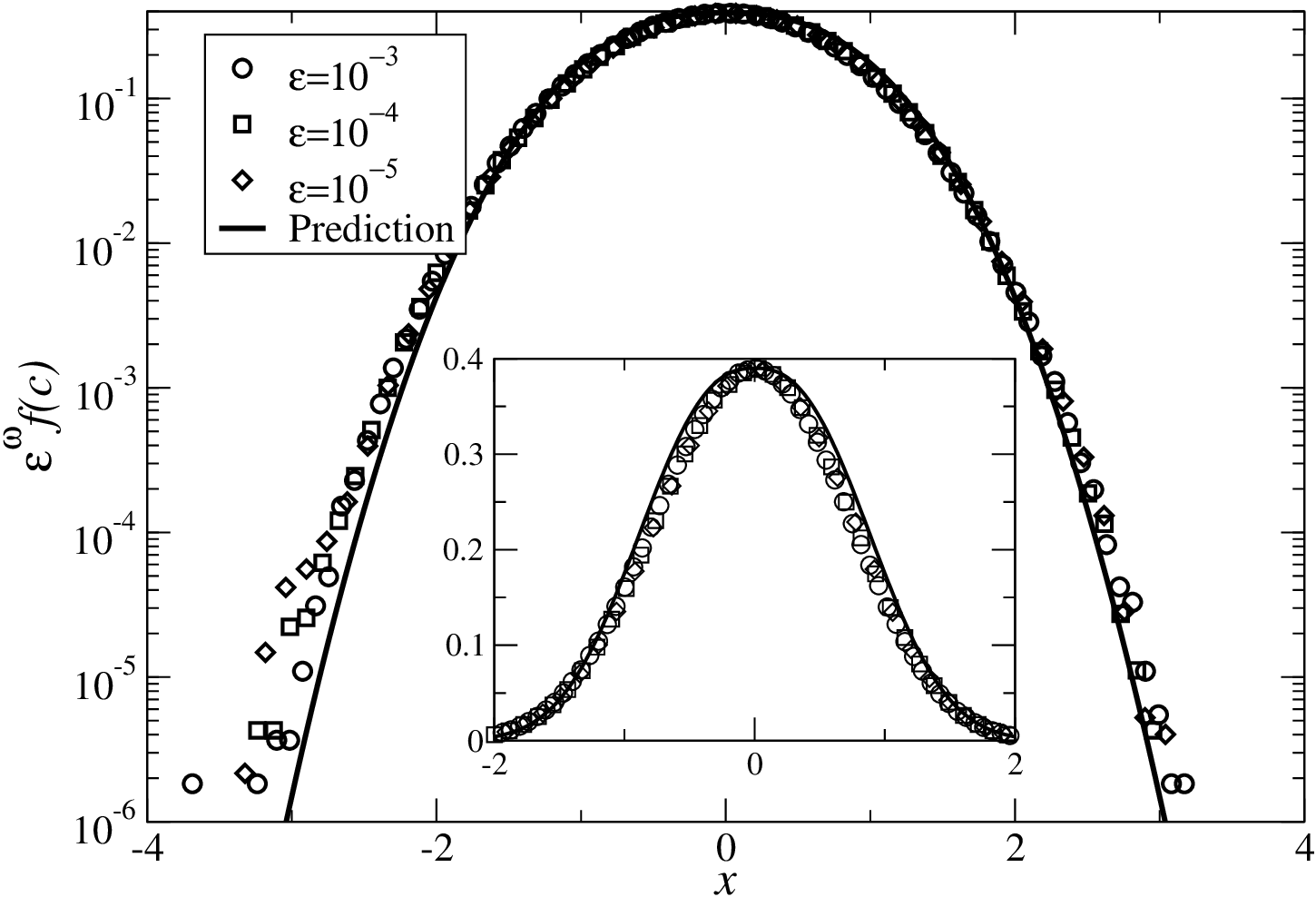}
\vskip -4.5mm \caption{ Scaling of type $\beta$. Same
plot as Figure \ref{fig:alpha}, displaying $\varepsilon^{\omega} f(c)$ vs.
$x= (|c|-c^*)/\varepsilon^{\omega}$ at $(\theta,\nu)=(3/4,1/2)$ with
$\omega=1/(\nu+2)=2/5$ and $c^* =1/\sqrt{2}$. The continuous curve shows
the prediction of Eq. (\ref{eq:psibeta}) where the prefactor is determined
from  the constraint $\int dc\psi=1$.} \label{fig:beta1}
\end{figure}

{\it Type $ \beta$-scaling:} On the line $\nu+1-2\theta=0$, the term of
order $q^{2\omega}$  in (\ref{eq:psi}) vanishes. If $ 0 < \nu <1$, the
terms $q$ and $q^{(\nu+2)\omega}$ can be balanced with the result
$\omega=1/(\nu+2)$, and one finds at large $|x|$,
\be
\psi(x) \propto \exp \left( -\frac{2}{\nu+2} |x|^{\nu+2}\right).
\label{eq:psibeta}
\ee
Unlike in $\alpha$-scaling , it is not possible here to obtain $\psi$ in
close form. We will refer to (\ref{eq:psibeta}) together with
$\omega=1/(\nu+2)$ as a scaling of type $\beta$. Figure \ref{fig:beta1}
shows that this behavior is in good agreement with the simulation results.
Note that  Eq. (\ref{eq:psibeta}) seems to hold also for small values of
$x$, whereas it is a priori only valid to describe the large$-x$ tail of
$\psi$. We also note that the large$-x$ tail (\ref{eq:psibeta}) exhibits
the same exponent $\nu+2$ as in White Noise driving (see Eq.
(\ref{f-asympt})). The important qualitative difference between the
velocity distributions reported in section \ref{sec:WN}  and here is: with
WN driving $f(c)$ is regular when $\alpha_0\to 1^-$, and with Negative
Friction it develops singularities.

On the line $\nu+1-2\theta=0$, now with $\nu>1$, one can only  balance the
terms in $q$ and $q^{3\omega}$ in (\ref{eq:psi}). This leads to
$\omega=1/3$, but the associated function $\psi$ diverges for large
arguments, and is thus unphysical. This is an indication that the double
peak limiting form cannot be valid  on the line $\nu+1-2\theta=0$ if
$\nu>1$. Monte Carlo results confirm this. They  display a limiting form
with three peaks in this region of parameter space (see Figure
\ref{fig:3peaks} and section \ref{sec:range} for a more thorough
discussion, in particular Figure \ref{fig:diagram}).

The special point on that line with $(\theta,\nu) =(1,1)$ represents free
cooling of inelastic hard rods, which has been studied in \cite{BBRTvW}.
There it was shown that (\ref{eq:ansatz}) holds, with $\omega=1/3$ and an
{\it asymmetric} scaling function, behaving at large arguments as,
\ba \label{psi-asympt}
\psi(x) & \simeq  C\exp\left[\frac{1}{3}x^3+o(1)\right] & \qquad
(x\to-\infty)
\nn \psi(x) &\simeq  C'\exp\left[-2 x\langle y^2\rangle_\psi \;
+o(1)\right]\quad \hbox{with}\quad
C'\,=\,C\,\exp\left[-\frac{1}{3}\langle y^3\rangle_\psi\right] &\qquad
 (x \to +\infty).
\ea
This asymmetry looks quite singular since most other scaling functions
identified so far are symmetric. However, more exceptional cases with
asymmetric scaling forms  $\psi$ can be found in Figure \ref{fig:nu-1}
(which corresponds to the point $(0,-1)$, a case of marginal stability
where $\nu=\theta-1$), as well as  in the scaling shapes of the ``Zoo''
region of Figure \ref{fig:diagram}.  We also emphasize that the Dirac
peaks appearing at the level of description when $\alpha_0\to 1^-$ are not
artifacts of discarding any spatial dependence in (\ref{eq:BE}), but
provide the exact solution of the Boltzmann equation where due account is
taken of the spatial degree of freedom of the particles. This has been
confirmed in \cite{Caglioti} and \cite{BBRTvW}, where the velocity
distributions of the homogeneous Boltzmann equation has been compared with
its exact counterpart, obtained by Molecular Dynamics simulations.

\subsubsection{Case $\nu=0$ ($\alpha'$-scaling).}

\begin{figure}[t]
\vskip .5cm
\includegraphics[height=5cm,angle=0]{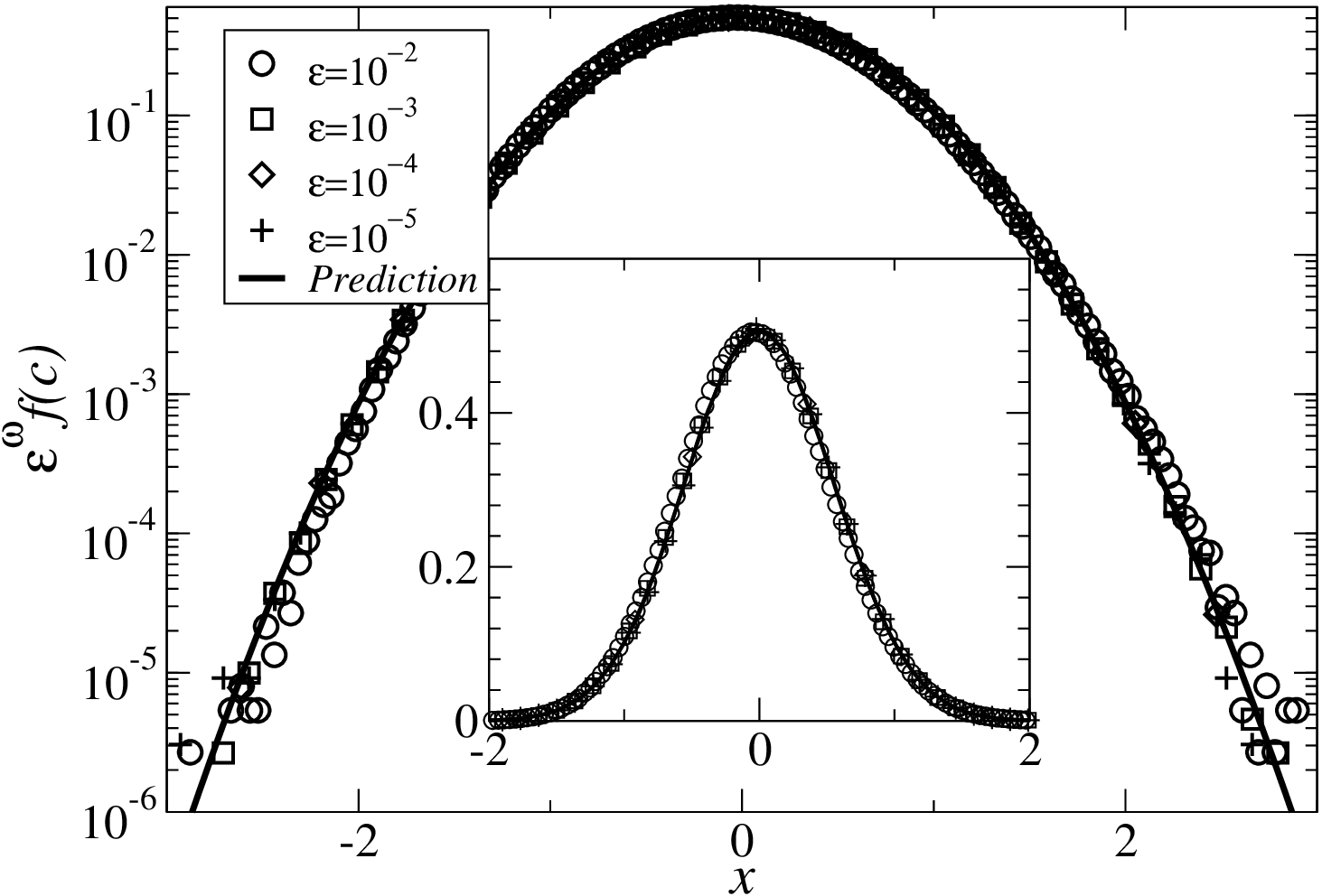}
\vskip -4.5mm \caption{ Scaling of type
$\alpha^\prime$. Same plot as Figure \ref{fig:alpha} and Figure
\ref{fig:beta1}, displaying $\varepsilon^{\omega} f(c)$ vs. $x=
(|c|-c^*)/\varepsilon^{\omega}$ at $(\theta,\nu)=(0.2, 0.0)$ with
$\omega=1/2$ and $c^* =1/\sqrt{2}$. The continuous curve is for Eq.
(\ref{eq:psialphap}), i.e. $\propto \exp [-(1-\theta) x^2)]$.}
\label{fig:alphaprime}
\end{figure}

{\it Type $\alpha^\prime$-scaling:} When $\nu$ vanishes (Maxwell models),
the first three terms on the rhs of (\ref{eq:psi}) are of the same order,
so that $\omega=1/2$ and $ \psi$ satisfies,
\be
\psi'(x) +2(1-2\theta)x\psi(x) + 2\psi(x)\langle x-y \rangle_\psi =
\psi'(x) + 4 (1-\theta) x \psi(x)=0,
\ee
since $\langle y \rangle_\psi =0$. Its solution is,
\be
\psi(x) \propto \exp [ -2(1-\theta) x^2]. \label{eq:psialphap}
\ee
This scaling, coined $\alpha'$, {\it a priori} holds for $0 \le \theta
\leq 1$, as follows from $\nu=0$ and the stability requirement
$\nu+1-\theta\geq0$. However, when $\theta=1$ (free cooling),
(\ref{eq:psialphap}) becomes unphysical. This is a consequence of the
peculiar behavior of the freely cooling one-dimensional Maxwell model: the
velocity distribution, which is algebraic, does not depend on $\alpha_0$
for $\alpha_0<1$ \cite{Maxwell,Balda}. In the (trivial) quasi-elastic
limit, $f$ can consequently not develop a singularity of any kind. The
scaling ansatz (\ref{eq:ansatz}) has to break down for $\theta=1$, and it
does. For $\theta<1$, the simulation data are in good agreement with
 $\alpha'$-scaling predictions (see Figure \ref{fig:alphaprime}), where again
no fitting parameter has been used.

\subsubsection{Case $\nu < 0 $ ($\beta$-scaling).}
{\it Type $\beta$-scaling:} Finally we  investigate the region $\nu <0$,
which is further confined by the stability requirements for the steady
state, $(\theta \geq 0, \nu \geq \theta -1 )$. We then have $(2+\nu)\omega
< 2\omega$ and the terms of order $q$ and $q^{(2+\nu)\omega}$ balance each
other in Eq. (\ref{eq:psi}). This implies $\omega=1/(\nu+2)$ and
\be
\frac{\psi'(x)}{\psi(x)}= -2\langle |x-y|^\nu (x-y)\rangle_\psi.
\ee
We recover the $\beta$-scaling,  and in particular the large$-|x|$
expression,
\be
\psi(x) \propto \exp \left( -\frac{2}{\nu+2} |x|^{\nu+2}\right).
\label{eq:psibeta2}
\ee
In the region $\nu<0$, we have successfully tested the validity of
$\beta$-scaling against Monte Carlo simulations. In some cases however, we
note that the best rescaling with respect to inelasticity is obtained with
an exponent $\omega_{\text{opt}}$ that slightly differs from  the
predicted $1/(\nu+2)$. For instance, with $(\theta,\nu) = (0.7,-0.2)$ we
find $\omega_{\text{opt}}\simeq 0.58$ while $1/(\nu+2)\simeq 0.55$. This
could indicate that the scaling limit has not yet been reached, or it
reflects the fact that for negative values of $\nu$, it is more difficult
to reach the steady state in Monte Carlo simulations. Here the collision
frequency is dominated by encounters with $|c_1-c_2|\ll 1$, that lead to a
negligible change in the velocities of colliding partners. Consequently
the numerical efficiency of our algorithm drops significantly.

\subsection{Range of validity of scaling predictions: towards a phase diagram.}
\label{sec:range}

We have reported above a good agreement between DSMC calculations and the
scaling predictions assuming the limiting double peak forms of
$\alpha$-,$\beta$- and $\alpha^\prime$-scaling, when $\varepsilon= 1-\alpha_0
\to 0^+$, for several points in the ($\theta,\nu$)-plane. We have also
shown that in some (complementary) regions of this plane the scaling
hypothesis (\ref{eq:ansatz}) does not provide an attracting fixed point
solution of the stationary nonlinear Boltzmann equation (\ref{eq:BE}) in
the quasi-elastic limit. In fact, our analysis shows that physical
solutions with $ \alpha$-scaling do {\it not} exist for $\nu \geq 0 $, and
$\nu+1 < 2\theta $ (e.g. $(\theta,\nu )= (1.1, 0.3); (1.1, 1.1)$), and
likewise  for $\beta$-scaling with $\nu >1$ and $ \nu +1 = 2 \theta$ (e.g.
$(\theta,\nu )= (1.1, 1.2) $). In these regions we have no predictions.

Furthermore, in the triangular region $[\theta >1,\nu >1, \nu+1> 2\theta]$
the NF driven kinetic equation admits -- at least on the basis of the
criteria developed -- QE-limiting solutions with two peaks, consistent at
least to second order in $ \varepsilon $. The dynamics selects a different
solution with three peaks  (see Figure \ref{fig:3peaks} with $(\theta,\nu
)= (1.1, 1.3)$).

\begin{figure}[thb]
\vskip .5cm
\includegraphics[height=5cm,angle=0]{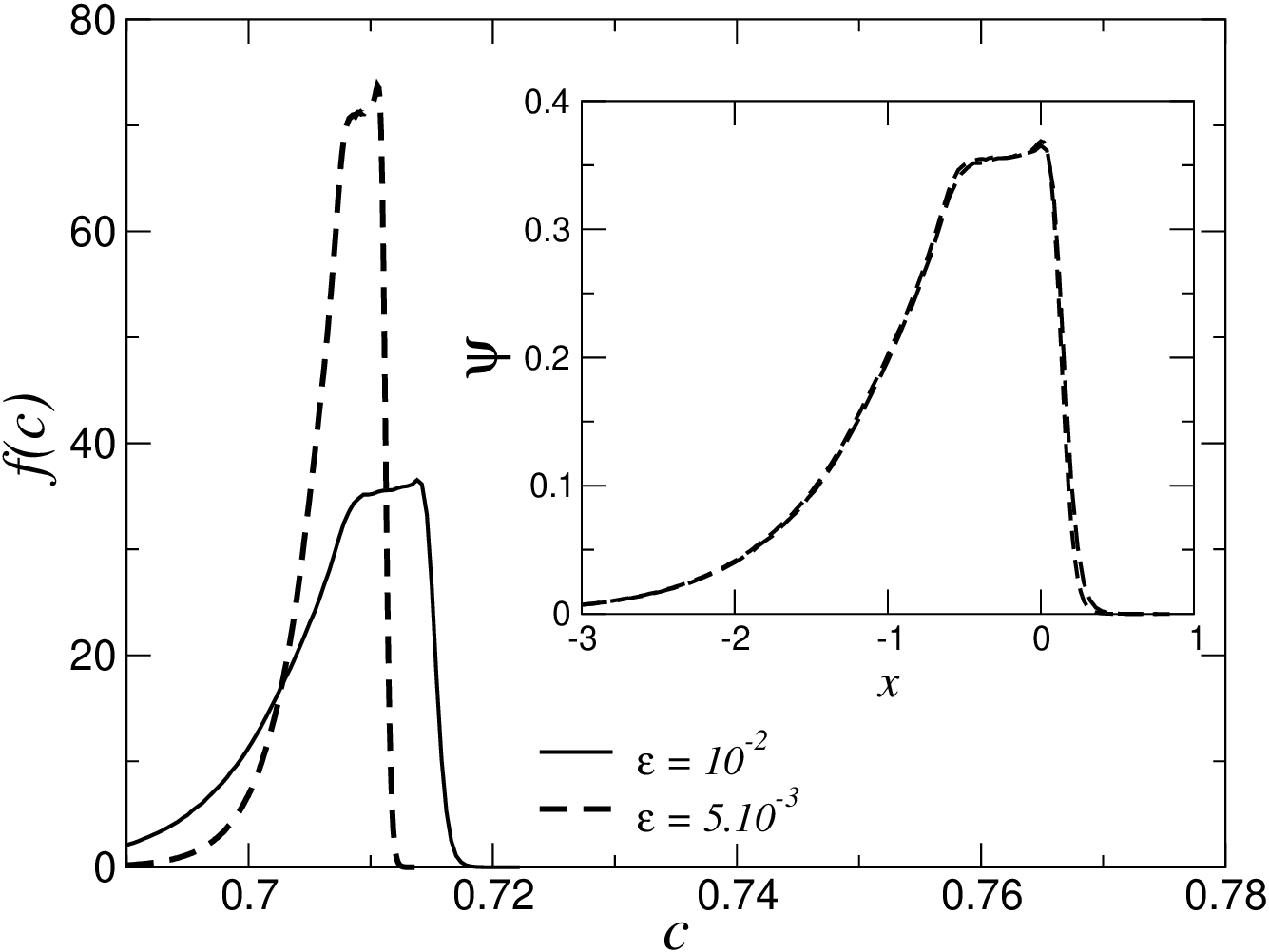}
\vskip -4.5mm \caption{Velocity
distribution at $(\theta,\nu) = (0,-1)$ for two inelasticities. $f(c)$
develops a cusp at $c=c^*$ where $c^*$ is $\varepsilon$-dependent. Plotting
$\psi(x)= \varepsilon f(c)$ vs. $x = (c-c^*)/\varepsilon$ (inset), shows that
here the scaling form \Ref{eq:ansatz} applies with an asymmetric $\psi$,
and $\omega=1=1/(\nu+2)$. For such low values of $\nu$, the CPU time
required to gather statistical knowledge is particularly large. }
\label{fig:nu-1}
\end{figure}

\begin{figure}[thb]
\vskip .5cm
\includegraphics[height=5cm,angle=0]{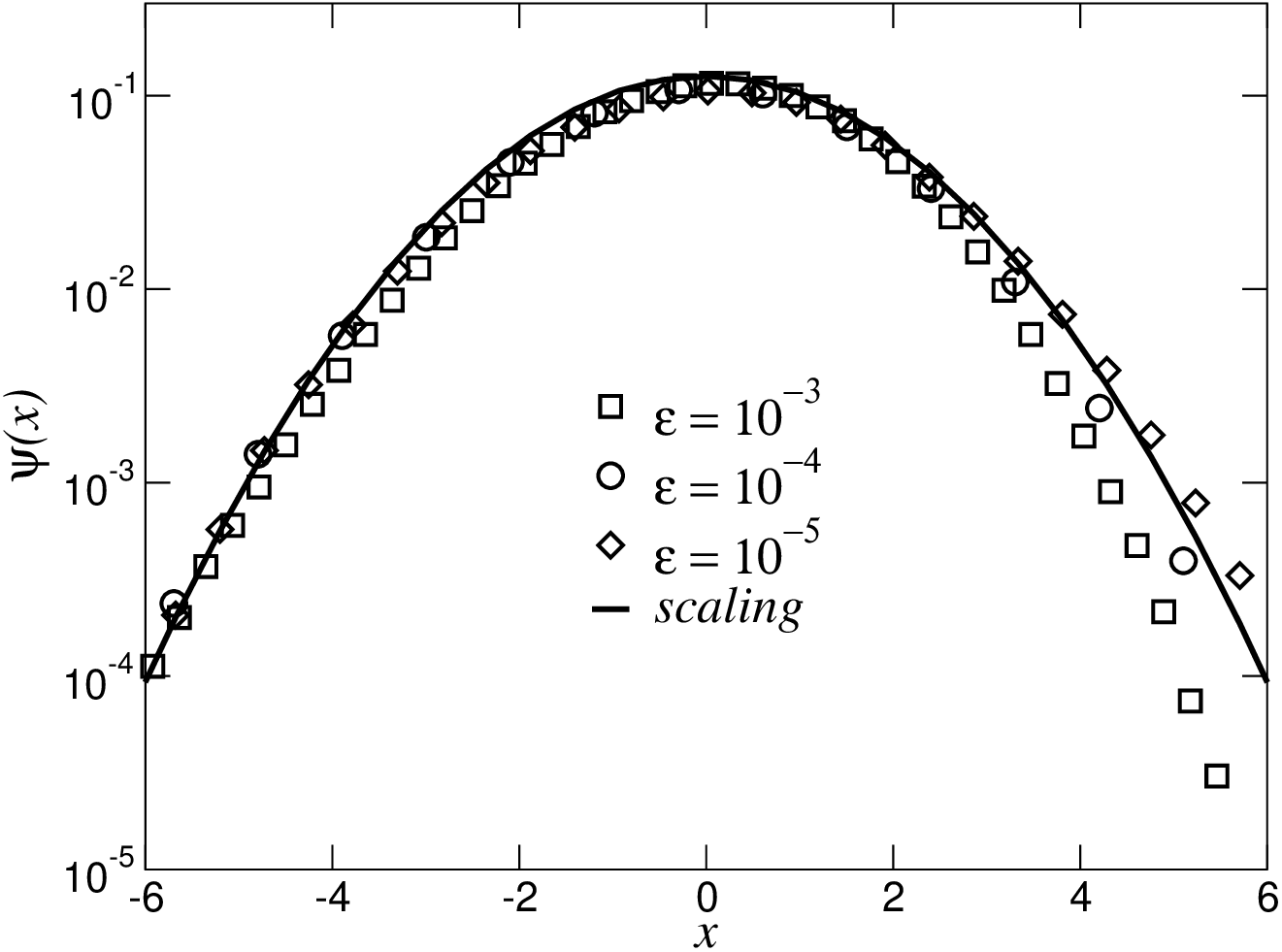}
\vskip -4.5mm \caption{Plots of
$\psi(x)=\varepsilon^{\omega} f(c)$ vs. $x = (c-c^*)/\varepsilon^{\omega}$, with
$\omega=1/2$ for the same parameters as in Figure~\ref{fig:3peaks}
($\theta, \nu) = (1.1, 1.3)$, i.e. for $f(c)$ approaching a structure
composed by three Dirac peaks. Here, $c^*\simeq 0.76$ denotes the position
of the right peak seen in Figure~\ref{fig:3peaks}. The continuous curve
shows expression (\ref{eq:psialpha}).} \label{fig:3peakspsi}
\end{figure}

\begin{figure}[t]
\vskip .5cm
\includegraphics[height=5cm,angle=0]{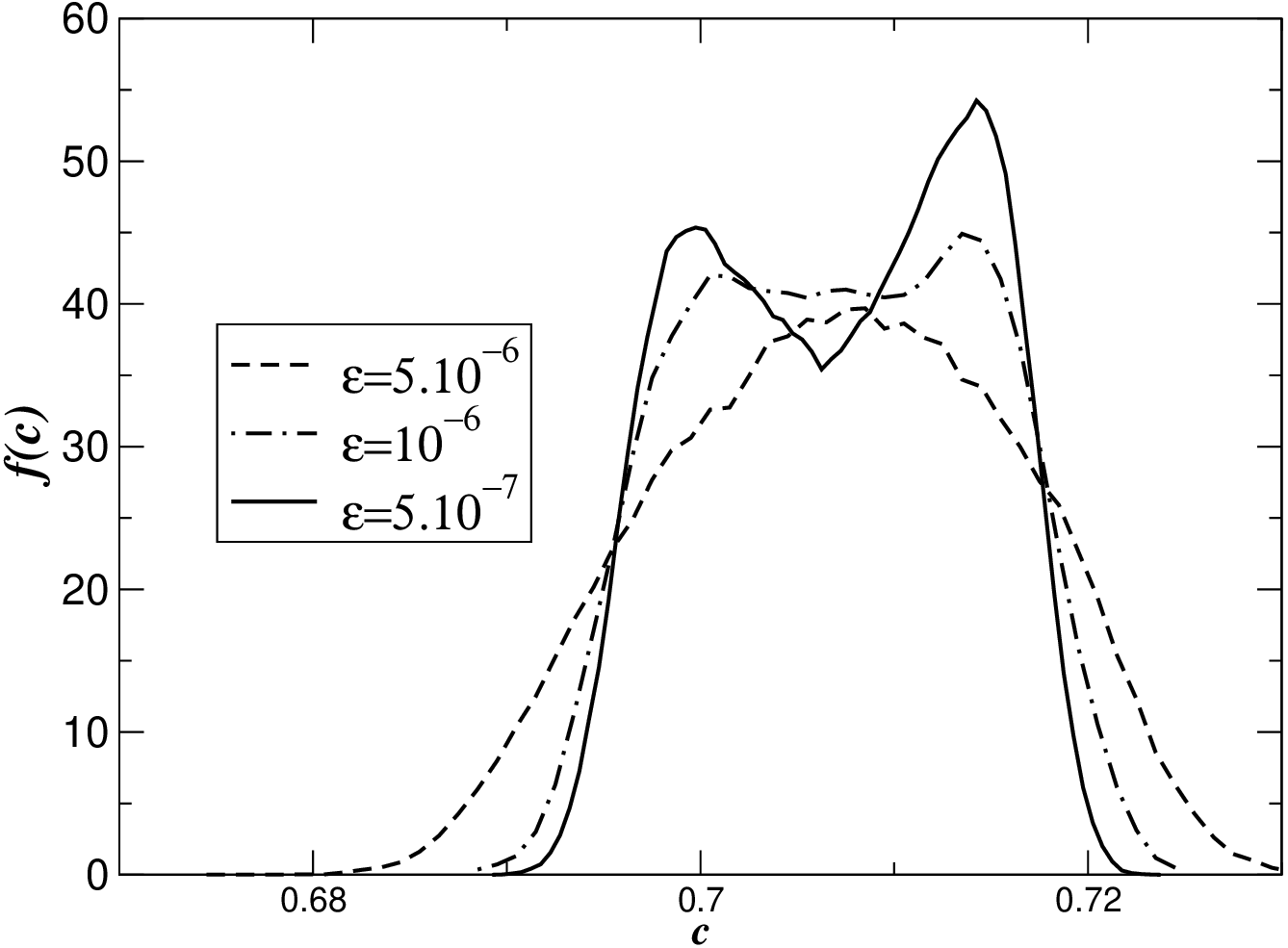}
\hskip .7cm
\includegraphics[height=5cm,angle=0]{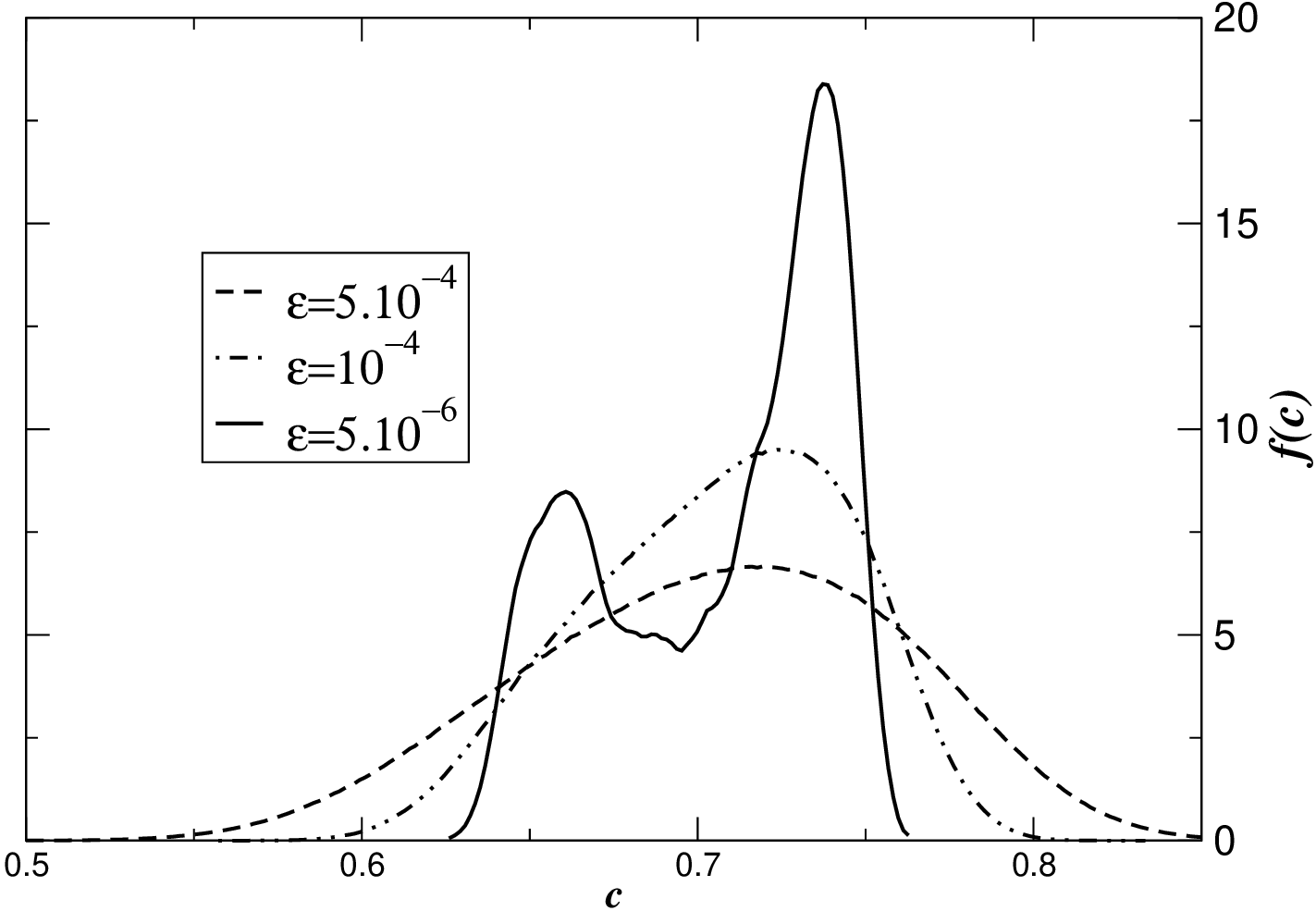}
\vskip -4.5mm \caption{Two Zoo members at $(\theta,\nu)=(0.75,0.3)$ (left) and
$(\theta,\nu)=(0.85,0.5)$ (right). Note the $x$- and $y$-scale in the plot
on the right. For these parameters one needs to decrease $\varepsilon$ below
$10^{-6}$ to realize that a seemingly single peak is likely to split into
two sub-peaks as $\varepsilon\to 0^+$. } \label{fig:Zoo1}
\end{figure}

\begin{figure}[t]
\vskip .5cm
\includegraphics[height=5cm,angle=0]{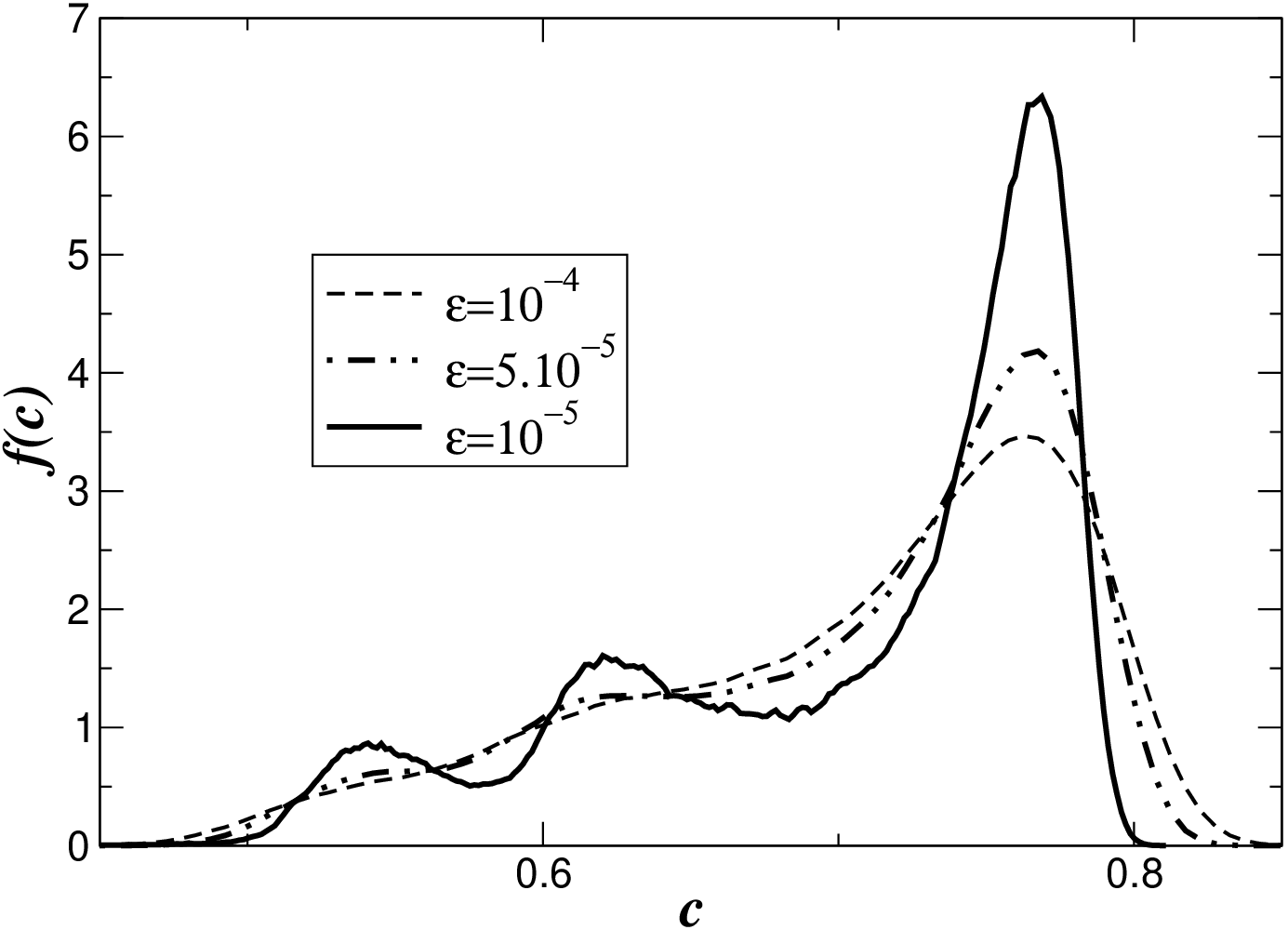}
\vskip -4.5mm \caption{ 
Another Zoo member at $(\theta,\nu)=(0.9,0.5)$. The velocity
distribution seems to evolve towards a six-peak form as $\varepsilon\to
0^+$.} \label{fig:Zoo2}
\end{figure}

A systematic Monte Carlo   investigation of various points ($\theta,\nu$)
--shown as dots in Figure \ref{fig:diagram}-- reveals that the range of
validity of $\alpha$-scaling  is in fact limited to $\theta\leq 1$
\cite{rque}.  In principle it would be possible to repeat the analysis of
section \ref{sec:scal2peaks} with a three-peak solution, however we did
not try to carry out this analysis. The reason is three-fold: First, it
would not explain why  $\alpha$-scaling is {\it not} selected by the
dynamics in the angular region above. Second, it is cumbersome. Third,
there is numerical evidence that $f(c)$ may differ from the two-peak or
three-peak form (see the Zoo region of Figure \ref{fig:diagram}). Thus,
such analysis would in any case not provide a complete picture. A
numerical investigation appears unavoidable and was used to identify the
different regions of the ``phase diagram'' shown in Figure
\ref{fig:diagram}. We summarize our main findings:
\begin{itemize}
\item $\alpha$-Scaling  holds for $\nu>0$ and $\nu>2\theta-1$ as found
analytically, with the restriction $\theta\leq 1$ that follows from
numerical evidence.
\item $\beta$-Scaling  applies to the line
$0<\nu=2\theta-1<1$ and also to the region $\nu<0$, where the additional
restriction $\nu > \theta-1$  follows from the stability requirement of
the steady-state solution of Eq. (\ref{eq:BE}). Figure \ref{fig:nu-1} with
$(\theta,\nu)= (0,-1)$ shows a case of marginal stability where
$\nu=\theta-1$. It is observed that the scaling exponent is still given by
$\omega=1/(\nu+2)$ but the form (\ref{eq:psibeta2}) breaks down ($\psi$
becomes asymmetric).
\item  $\alpha'$-Scaling is valid  on the ``Maxwell'' line $\nu=0$ for $\theta < 1$.
\item Hard rods under free cooling, i.e. $(\theta,\nu)=(1,1)$,  display specific
scaling (see  Eq. \Ref{psi-asympt}, Table \ref{table:table} and Ref.
\cite{BBRTvW} for details).
\item None of the above scalings hold for $\theta>1$, where we have always
observed a triple peak as in Figure \ref{fig:3peaks}. Figure
\ref{fig:3peakspsi} (with the same parameters as Figure \ref{fig:3peaks})
shows that the distributions are also self-similar, with an exponent
$\omega=1/2$. Although, we have no prediction for the three-peak forms, we
note that the scaling function in Figure \ref{fig:3peakspsi} is compatible
with Eq. (\ref{eq:psialpha}), pertaining to $\alpha$-scaling. This could
be specific to the parameters chosen, since those value of $\theta$ and
$\nu$ obey the inequalities, $\nu>0$ and $\nu>2\theta-1$, where
$\alpha$-scaling provides a two-peak solution of the Boltzmann equation.
Two- and three-peak shapes therefore seem to have common features. We did
not explore self-similarity further in the three-peak region.
\item There exists another triangular region in the ($\theta,\nu$)-plane
(the Zoo in Figure \ref{fig:diagram}), where $f(c)$ does not evolve toward
a two-peak or a three-peak form. In some instances, the Monte Carlo data
are compatible with a four-peak limit as $\varepsilon \to 0^+$ (see Figure
\ref{fig:Zoo1} where only the  sector with $c>0$ has been shown). In some
other cases, we observe precursors of what seems to be a six-peak form
(see Figure \ref{fig:Zoo2}).  For both  Figures \ref{fig:Zoo1} and
\ref{fig:Zoo2}, it is difficult to decide if the limiting form for
$\varepsilon\to 0^+$ will be a collection of Dirac distributions (with
therefore a support of vanishing measure), or a distribution with finite
support. We could however identify some points in the triangle where the
limiting $f(c)$ clearly is of finite support (see Figure
\ref{fig:complex}).
\end{itemize}
To summarize,  $\alpha'$- and $\beta$-scaling apply in the whole domain
where they provide a solution to the Boltzmann equation, but
$\alpha$-scaling has a restricted domain of relevance compared to the
region where the corresponding solution is self-consistent. The key
features of the analytical predictions are recalled in Table
\ref{table:table}.

\begin{table}[thb]
\begin{tabular}{|l|c|c|c|}
\hline
Scaling type & rescaling exponent & rescaling function \\
\hline
\hline
$\alpha$ & $ \omega=1/2$  & $\psi(x) \propto \exp (-(\nu+1-2\theta) x^2)$\\
\hline
$\alpha'$ &   $\omega=1/2$ &
$\psi(x) \propto \exp (-2(1-\theta) x^2)$ \\
\hline
$\beta$ &   $\omega=1/(\nu+2) $ &
~~$\psi(x) \sim \exp (-2 |x|^{\nu+2} /(\nu+2))$ at large $|x|$~~ \\
\hline
$\theta=1,\nu=1$ &$\omega=1/3$ & Eq. (\ref{psi-asympt}), $\psi$ asymmetric \\
\hline
\end{tabular}
\caption{Theoretical predictions for the different scaling behaviors in
the two-peak region.}
\label{table:table}
\end{table}

\section{Conclusion}

We have studied the one-dimensional nonlinear Boltzmann equation in the
limit of quasi-elastic collisions for a class of dissipative fluids where
material properties are encoded in an exponent $\nu$ such that the
collision frequency between two particles with velocities $c_1$ and $c_2$
scales like $|c_1-c_2|^\nu$.  Two driving mechanisms have been considered:
stochastic {\it white noise} (in which case the generic effects only
depend on $\nu$) and deterministic {\it negative friction} (in which case,
in addition to $\nu$, a second important exponent $\theta \geq 0$,  
characterizing the driving, has been
considered). In both stochastic and deterministic cases, the quasi-elastic
limit does not commute with the limit of large velocities. This is
specific to one space dimension.  There are however important differences
between the two driving mechanisms. In the white noise case, the
normalized velocity distribution $f(c)$, suitably rescaled to have fixed
variance, is regular and displays stretched exponential QE-tails of the
form $\exp[-c^{\nu+2}]$. On the other hand, $f(c)$ with deterministic
driving --which encompasses the much studied homogeneous cooling regime--
develops singularities as $\varepsilon\equiv1-\alpha_0 \to 0^+$.  The
corresponding scaling behavior is particularly rich.  We have classified
the scaling forms encountered in several families, see Figure
\ref{fig:diagram} for a global picture.  Some regions of this
$(\theta,\nu)$-diagram are well understood, such as regions $\alpha$,
$\alpha'$ and $\beta$.  Some other domains resist theoretical
understanding. Even if some progress might be possible in the three-peak
region (one at $c=0$, the two others at $\pm c^*$), the situation in the
central Zoo region of Figure \ref{fig:diagram} seems more difficult to
rationalize, and computationally elusive, since one needs to reach
extremely small values of $\varepsilon$ to see the precursors of presumed
singularities.

\begin{appendix}
\section{}
In this appendix, we expand the collision operator $I(c|f)$ defined in
(\ref{NL-coll1}) in powers of $q=(1-\alpha)/2$, up to second order. Such
an expansion is required to unveil the internal structure of the singular
peaks that develop as $q\to 0^+$ with driving by negative friction.
Assuming that the functional form of the velocity distribution is given by
(\ref{eq:ansatz}), our goal is to obtain here the differential equation
fulfilled by the scaling function $\psi$. Starting from
\be
I(c|f)= \int du |c-u|^\nu f(u) \left[
\frac{1}{p^{\nu+1}} f\left(c+\frac{q}{p}(c-u)\right) -f(c)\right],
\ee
one obtains by extending \Ref{QE-coll} to ${\cal O}(q^2)$ included,
\be
I(c|f)= q \frac{d}{dc} \left[
f(c) \int du f(u) |c-u|^\nu (c-u) \right]
+\frac{q^2}{2} \left(\frac{d}{dc} \right)^2 \left[ f(c)
\int du |c-u|^{\nu+2} f(u) \right] + {\cal O}(q^3).
\label{eq:a2}
\ee
Inserting the ansatz (\ref{eq:ansatz}) into (\ref{eq:a2}), the term
$I^{(1)}$ of order $q$ reads
\ba
I^{(1)} = -q \left(\frac{b}{4E}\right)^2 \left(\frac{2}{b}\right)^{\nu+1}
\frac{d}{dx} \left\{ \psi(x) \int dy \psi(y) (1-E(x+y))^{\nu+1} \right\}
\nonumber \\
+q \left(\frac{b}{4E}\right)^2 \left(\frac{2E}{b}\right)^{\nu+1}
\frac{d}{dx} \left\{ \psi(x) \int dy \psi(y)|x-y|^\nu (x-y) \right\},
\ea
and expanding  $ (1-E(x+y))^{\nu+1}$ further yields,
\be
I^{(1)} =  \frac{q}{4E^2} \left(\frac{2}{b}\right)^{\nu-1} \frac{d}{dx}
\left\{\left[ -1 + xE(\nu+1) -\frac{\nu(\nu+1)}{2} E^2 \int dy \psi(y)
(x+y)^2 +E^{\nu+1} \int dy \psi(y)|x-y|^\nu (x-y)  \right]\psi(x)\right\}.
\ee

The second order term $ I^{(2)} $ in (\ref{eq:a2}) is
\be
I^{(2)} = \frac{q^2}{2} \left(\frac{b}{4E}\right)^2 \frac{b}{2E}
\left(\frac{d}{dx}\right)^2 \left\{
\left(\frac{2E}{b}\right)^{\nu+2}\psi(x) \int dy \psi(y) |x-y|^{\nu+2}
+\left(\frac{2}{b}\right)^{\nu+2}\psi(x) \int dy \psi(y)
|1-E(x+y)|^{\nu+2} \right\}
\ee
and we can keep only the largest term as $q \to 0^+$,  i.e.
\be
\frac{q^2}{8E^3} \left(\frac{2}{b}\right)^{\nu-1} \psi''(x).
\ee
Collecting terms yields finally,
\ba
I(c|f)=
\frac{q}{4E^2} \left(\frac{2}{b}\right)^{\nu-1}
\frac{d}{dx}
\left\{\left[-1+ xE(\nu+1)
-\frac{\nu(\nu+1)}{2} E^2 \int dy \psi(y) (x+y)^2 \right.\right. \nonumber \\
\left.\left. +E^{\nu+1} \int dy \psi(y)|x-y|^\nu (x-y)  \right]\psi(x)
+\frac{q}{2E}\psi'(x) \right\}.
\label{eq:a7}
\ea
The next step is to evaluate the right hand side of equation
(\ref{BE-NF}), by an expansion of the moments $\kappa_\nu$ and
$\mu_{\theta+1}$,
\ba
\mu_{\theta+1}&=& \int dc |c|^{\theta+1} f(c)
= \int dy \psi(y) \left| \frac{1-2Ey}{b} \right|^{\theta+1} \nonumber \\
&=& b^{-1-\theta} [1+2\theta(\theta+1)E^2 \langle y^2 \rangle_\psi]
\ea
\ba
\kappa_\nu&=& \int\int dc du |c-u|^{\nu+2} f(c) f(u)\\
&=& \frac{2^{\nu+1}}{b^{\nu+2}} \left[ \int dy dz \psi(y)\psi(z)
(1-E(y+z))^{\nu+2}
+E^{\nu+2} \int dy dz \psi(y)\psi(z) |y-z|^{\nu+2} \right]\nonumber \\
&=& \frac{2^{\nu+1}}{b^{\nu+2}} \left[1+(\nu+1)(\nu+2) E^2 \langle y^2
\rangle_\psi +E^{\nu+2} \int dy dz \psi(y)\psi(z) |y-z|^{\nu+2} \right].
\ea
With $c$ close to $-1/b$, one can also perform the expansion
\be
\frac{d}{dc} \left( c |c|^{\theta-1} f \right) = -
\frac{b^{2-\theta}}{8E^2} \frac{d}{dx} \left\{ \psi(x) [1-2\theta E x + 2
\theta(\theta-1)E^2x^2], \right\}
\ee
and the rhs of equation (\ref{eq:ansatz}) can thus finally be written as
\ba
&pq \frac{\kappa_{\nu}}{2 \mu_{\theta+1}}
 \frac{d}{dc} \left( c |c|^{\theta-1} f \right)= -
\frac{pq}{4E^2} \left( \frac{2}{b} \right)^{\nu-1}
\frac{d}{dx} \left\{
\psi(x) [1-2\theta E x +
2 \theta(\theta-1)E^2x^2
\right.\nonumber \\
&+ \left. ((\nu+1)(\nu+2)-2\theta(\theta+1)) E^2 \langle y^2 \rangle_\psi
+ E^{\nu+2} \int \int dy dz |y-z|^{\nu+2} \psi(y)\psi(z) ] \right\}&.
\ea
We obtain (\ref{eq:psi}), after
simplification by the prefactors and integration
with respect to $x$.
\end{appendix}



\begin{thebibliography}{99}

\bibitem{Gold}
I. Goldhirsch, Ann. Rev. Fluid Mech. {\bf 35}, 267 (2003).

\bibitem{PoschelBrill}
{\it Theory of Granular Gas Dynamics}, Th. P\"oschel and N.V.
Brilliantov (Eds), (Springer-Verlag, Berlin, 2003).

\bibitem{usJPCM}
A. Barrat,  E. Trizac,  M.H. Ernst,
J. Phys. Condens. Matter {\bf 17}, S2429 (2005).

\bibitem{Mcnamara2}
S. McNamara and W.R. Young, Phys. Fluids A {\bf 5}, 34 (1993).

\bibitem{Caglioti}
D. Benedetto, E. Caglioti and M. Pulvirenti, Math. Mod. and Num. An.
{\bf 31}, 615 (1997).

\bibitem{Benedetto}
D. Benedetto, E. Caglioti, J.A. Carrillo and M. Pulvirenti,
J. Stat. Phys. {\bf 91}, 979 (1998).

\bibitem{Ramirez}
R. Ramirez and P. Cordero, Phys. Rev. E {\bf 59}, 656; also in
{\it Granular Gases}, eds T. Poschel and S. Luding (Springer, NY, 2001).

\bibitem{Mcnamara1}
S. McNamara and W.R. Young, Phys. Fluids A {\bf 4}, 496 (1992).

\bibitem{Sela}
N. Sela and I. Goldhirsch, Phys. Fluids {\bf 7}, 507 (1995).

\bibitem{bennaim}
E. Ben-Naim, S.Y. Chen, G.D. Doolen and S. Redner, Phys. Rev. Lett. {\bf
83}, 4069 (1999).

\bibitem{Rosas}
A. Rosas, D. ben-Avraham and K. Lindenberg,
Phys. Rev. E {\bf 71}, 032301 (2005).

\bibitem{EB-Rap02} M.H. Ernst and R. Brito, Phys. Rev. E \textbf{65}, 040301(R)
(2002).

\bibitem{ETB}
M.H. Ernst, E. Trizac and A. Barrat, J. Stat. Phys. {\bf 124}, 549 (2006).

\bibitem{ETB-EPL06} M.H. Ernst, E. Trizac and A. Barat, Europhys. Lett. 
{\bf 76}, 56 (2006).

\bibitem{WN}
D.R.M. Williams and F.C.  MacKintosh, Phys. Rev. E {\bf 54}, R9 (1996); G.
Peng  and T. Ohta, Phys. Rev. E {\bf 58}, 4737 (1998); T.P.C. van Noije
and M.H. Ernst, Granular Matter {\bf 1}, 57 (1998); C. Henrique, G.
Batrouni and D. Bideau, Phys. Rev. E {\bf 63} 011304 (2000); 
S.J. Moon,
M.D. Shattuck and J.B. Swift, Phys. Rev. E {\bf 64} 031303 (2001); 
I. Pagonabarraga, E. Trizac, T.P.C. van Noije and M.H. Ernst, 
Phys. Rev. E {\bf 65}, 011303 (2002);
J.S. van Zon and F.C. MacKintosh, Phys. Rev. E {\bf 72}, 051301 (2005).

\bibitem{MS}
J.M. Montanero and A. Santos, Granular Matter \textbf{2}, 53
(2000).

\bibitem{BBRTvW}
A. Barrat, T. Biben, Z. R\`acz, E. Trizac and F. van Wijland, J. Phys. A:
Math. Gen. \textbf{35}, 463 (2002).

\bibitem{SE-PRE03} A Santos and M.H. Ernst, Phys. Rev. E \textbf{68}, 011305
(2003).

\bibitem{ME} M.H. Ernst, Physics. Reports {\bf 78}, 1 (1981).

\bibitem{Maxwell}
E. Ben-Naim and P.L. Krapivsky,
 Phys. Rev. E \textbf{61}, R5 (2000);
E. Ben-Naim and P.L. Krapivsky,
Phys. Rev. E \textbf{66}, 1309 (2002).

\bibitem{Balda}
A. Baldassarri, U.M.B. Marconi, and A. Puglisi,
 Europhys. Lett. \textbf{58}, 14 (2002);
A. Baldassarri, U.M.B. Marconi, and A. Puglisi,
Math. Mod. Meth. Appl. S. {\bf 12}, 965 (2002).

\bibitem{EBri}
M.H. Ernst and R. Brito,
J. Stat. Phys. {\bf 109}, 407 (2002).

 \bibitem{Bird}
G. Bird,
Molecular Gas Dynamics and the Direct Simulation of Gas Flows
(Clarendon, Oxford, 1994).

\bibitem{rque}
In this respect, Figure \ref{fig:alpha} corresponds to the frontier of the
domain of validity of $\alpha$-scaling. We have checked however that cases
with $\theta<1$ and $\nu>0$ belong to the $\alpha$ family, provided that
$\nu+1-2\theta>0$.

\end{thebibliography}
\end{document}